\documentclass[aps,prd,amsmath,amssymb,superscriptaddress,showkeys,showpacs,twocolumn,floatfix]{revtex4-1}

\immediate\write18{texcount -inc -sum 20260427_main.tex > wordcount.txt}

\usepackage{amsmath,amssymb,mathtools}
\usepackage{mathrsfs} 
\usepackage{graphicx,color}
\usepackage{float}
\usepackage{appendix}
\usepackage{hyperref}
\usepackage{verbatim}
\usepackage{comment}
\usepackage{subcaption}
\usepackage{bm}
\usepackage{xcolor}
\usepackage[justification=raggedright,singlelinecheck=false]{caption}

\usepackage[utf8]{inputenc}
\usepackage[english]{babel}
\usepackage{booktabs}
\usepackage[normalem]{ulem}

\begin{document}
\title{Enhancing Gravitational Waves from Global Strings via a First-Order Phase Transition}
\author{Yongtao Jia}
\affiliation{International Centre for Theoretical Physics Asia-Pacific (ICTP-AP), University of Chinese Academy of Sciences (UCAS), Beijing, China.}
\affiliation{Taiji Laboratory for Gravitational Wave Universe, University of Chinese Academy of Sciences (UCAS), Beijing, China.} 

\author{Ligong Bian\thanks{Corresponding Author.}}
\email{lgbycl@cqu.edu.cn}
\affiliation{~Department of Physics and Chongqing Key Laboratory for Strongly Coupled Physics, Chongqing University, Chongqing 401331, P. R. China}

\author{Huai-Ke Guo}
\email{guohuaike@ucas.ac.cn}
\affiliation{International Centre for Theoretical Physics Asia-Pacific (ICTP-AP), University of Chinese Academy of Sciences (UCAS), Beijing, China.}
\affiliation{Taiji Laboratory for Gravitational Wave Universe, University of Chinese Academy of Sciences (UCAS), Beijing, China.}

\begin{abstract}
We perform the first (3+1)-dimensional lattice simulations of a cosmological first-order phase transition (PT) in which true-vacuum bubbles with random phases nucleate, expand, and collide, producing a network of global strings. We find that the string network enters the scaling regime, and the coefficient of the logarithmic growth is smaller than the value obtained for a second-order PT. From the axion spectrum, we derive the dark-matter relic abundance and the dark-radiation bound from big bang nucleosynthesis on the axion decay constant $f_a$ and mass $m_a$. The ratio of the gravitational wave emission rate to the axion emission rate is found to be about two orders of magnitude larger than in the second-order case, and the resulting gravitational wave spectrum can be probed by pulsar timing arrays as well as by space- and ground-based interferometers.   
\end{abstract}
\maketitle
\section{Introduction}

The pulsar timing array (PTA) collaborations have reported possible evidence for a stochastic gravitational-wave (GW) background in the nHz frequency band ~\cite{NANOGrav:2023gor,EPTA:2023fyk,Reardon:2023gzh,Xu:2023wog}, offering a potential insight into detecting early-universe processes. 
In this context, cosmic strings, which are topological defects formed by the spontaneous breaking of a $U(1)$ symmetry~\cite{Kibble:1976sj,Vilenkin:1982ks}, can provide a viable explanation for the reported PTA signal~\cite{Kitajima:2023vre,Bian:2023dnv,Basilakos:2023xof,Wang:2023len,Ellis:2023tsl,Buchmuller:2023aus,Figueroa:2023zhu,NANOGrav:2023hvm,EPTA:2023xxk,Cheng:2024axj,Bian:2020urb}. Moreover, the axion emitted during cosmic string evolution, as a pseudo-Nambu-Goldstone boson arising from the breaking of the Peccei-Quinn (PQ) symmetry invoked to solve the strong CP problem~\cite{Peccei:1977hh,Peccei:1977ur,Weinberg:1977ma,Wilczek:1977pj,Kim:1979if,Shifman:1979if,Zhitnitsky:1980tq,Dine:1981rt}, is also an attractive dark matter candidate~\cite{Preskill:1982cy,Abbott:1982af,Dine:1982ah}. 

Both axion and GW emissions are essential ingredients in studies of string-network evolution for axion strings \cite{Auclair:2019jip,Matsunami:2019fss}. The scaling property and free axion emission in the string network evolution have been studied with the method of lattice field simulation for the second order PT process in recent years~\cite{Gorghetto_2018,Gorghetto:2020qws,Buschmann_2022,Benabou:2023ghl,Benabou:2024msj,Saikawa:2024bta,Kim:2024wku}. The GW emission has been investigated for string loops~\cite{Baeza-Ballesteros:2023say} and for the full string network~\cite{Vachaspati:1984gt,Gorghetto:2021fsn,Zhao:2025kvm}, with a characteristic suppression factor of $f_a^2/m^2_{pl}$.
Apart from the second-order PT predicted by the Kibble mechanism~\cite{Hindmarsh:1994re,Kibble:1980mv},  
it was proposed that the first-order PT can also produce the cosmic string through the geodesic rule when bubbles with random phases collide with each other~\cite{Hindmarsh:1993av,Copeland:1996jz,Kibble:1995aa}. 
Early simulations in (2+1)-dimensional spacetime investigated vortex formation in first-order PTs~\cite{Srivastava:1991nv, Melfo:1994cv, Ferrera:1996hu, Lilley:2001df}, focusing on the string production rate from bubble collisions and its dependence on parameters such as the bubble wall velocity.

This paper performs the first (3+1)-dimensional lattice simulations to study the dynamics of global strings produced by bubble collisions, together with the emission of free axions and GWs from these strings. The log dependence in the scaling parameter is confirmed to ensure that the network enters the scaling stage.
The axion decay constant $f_a$ and mass $m_a$ are subject to constraints from the dark matter relic density and dark radiation in Big Bang nucleosynthesis (BBN). We find that the ratio $\gamma$ of GWs energy to free axion energy is two orders of magnitude larger than in a second-order PT, which allows us to assess the detectability of GWs from global strings in GW detectors. 

\section{The simulation setup}
The action of the system under study is given by 
\begin{align}
    S = - \int \mathrm{d}^4 x (\partial_\mu \varphi \partial^\mu \varphi^*  + V(|\varphi|) \; ,
\end{align}
with the FLRW metric convention of $ ds^2 =a^2(\tau)(-d\tau^2+d\mathbf{x}^2) $, and the potential form is  
\begin{align}
V(|\varphi|)=\frac{1}{2} M^2 |\varphi|^2+\frac{1}{3} \delta |\varphi|^3+\frac{1}{4} \lambda |\varphi|^4
\end{align} 
with $\delta<0$ to trigger a first-order PT. The radial mode and the phase mode is defined as $\varphi = \frac{(r+f_a)}{\sqrt{2}} e^{i a / f_a }$, with $f_a = \sqrt{2}\phi_v$ and $\phi_v$ is the true vacuum value of the potential. The squared mass of the radial mode is $m_r^2 = M^2/2+\delta \phi_v+3\lambda\phi_v^2/2$. 

The simulation adopts a fixed-background expansion with the initial scale factor $a_i = 1.0$. The field and spacetime variables are rescaled as $\tilde{\varphi} = \varphi/f_\star$, $\delta\tilde{x} = \omega_\star\,\delta x$, and $\delta \tilde{\tau} = \omega_\star \, \delta t / a$, where $f_\star = \omega_\star = H_i$. 
The dimensionless potential form is 
$\tilde{V} = V/f_\star^2 \omega^2_\star = \frac{1}{2} \tilde{M}^2 |\tilde{\varphi}|^2+\frac{1}{3} \tilde{\delta} |\tilde{\varphi}|^3+\frac{1}{4} \tilde{\lambda} |\tilde{\varphi}|^4 $. By varying the action, the equation of motion in dimensionless form is obtained as
\begin{align}
        \tilde{\varphi}^{\prime \prime} - \tilde{\nabla}^2 \tilde{\varphi} + 2\frac{a^{\prime}}{a} \tilde{\varphi}^{\prime} &= -a^2\frac{\partial \tilde{V}}{\partial|\tilde{\varphi}|}\frac{\tilde{\varphi}}{2|\tilde{\varphi}|} \; , 
        \label{eq:eom}
\end{align}
 where the prime denotes the time derivative with respect to conformal time $\tilde{\tau}$. With the choice of $\omega_\star$ and $a_i$, the expansion in the radiation-dominated background in the simulation is given by 
\begin{equation}
    \tilde{\tau} = \frac{\tau}{\tau_i} = \frac{a}{a_i} = \left(\frac{T}{T_i}\right)^{-1} = \left(\frac{H}{H_i}\right)^{-\frac{1}{2}} \;.
    \label{eq:back}
\end{equation}
from the Friedmann equation. To simplify the simulation, the initial Hubble parameter is chosen to be $H_i = 7.11 \times 10^{17}\,\mathrm{GeV}$, which equals the initial temperature $T_i = H_i$ for a relativistic degree of freedom $g_\star = 106.75$. 

Using the public code ${\mathcal CosmoLattice}$~\cite{Figueroa:2021yhd,Figueroa:2020rrl}, we evolve the equation of motion Eq.~\ref{eq:eom} in the radiation-dominated background with a second-order leap-frog algorithm. Initially, bubbles nucleate with randomly distributed centers. The magnitude of the complex field follows the radial profile
\begin{align}
    |\varphi| = \frac{\phi_b}{2} (1-\tanh{\frac{R-R_0}{l_w}}) 
\end{align}
for $R<R_0+l_w/2$, where $R$ is the distance from the bubble center. Each bubble is assigned a single random phase uniformly distributed in $[0, 2\pi]$. The bubble profile is computed using \textbf{FindBounce}~\cite{Guada:2018jek,Guada:2020xnz}. Outside the bubble regions, the field values are set to zero. The simulation box consists of $N=1024$ grid points, and the rescaled time step is $\delta \tilde{\tau} = 0.01 < \delta \tilde{x} / 3 $ for each run. At the end of the simulation, the string core $d_s = m_r^{-1}$ is resolved by 1.4 grid spacings. 
To investigate the effect of the number of bubbles, we consider two parameter setups listed in Table.~\ref{tab:para}, where the tilde denotes dimensionless variables rescaled by $f_\star$ and $\omega_\star$. To investigate the randomness of the bubble places in the simulation box, each parameter setup is realized five times.

\begin{table}[!htp]
    \centering
    \begin{tabular}{c|ccccccccc}
    & $\tilde{M}$  & $\tilde{\delta}$ & $\tilde{\lambda}$ & $\tilde{\varphi}_b$ & $\tilde{R}_0$ & $\tilde{l}_w$ & $N_b$ & $\tilde{L}$ & $L/H_{e}^{-1}$ \\
    \hline     
    setup 1 & 0.64 & -3.32 & 2.37 & 0.91 & 2.48 & 1.91 & 238 & 48.07 & 4.18 \\
    setup 2 & 0.77 & -3.32 & 2.37 & 1.08 & 3.14 & 1.84 & 119 & 48.07 & 3.85 \\
    \end{tabular}
    \caption{The rescaled potential parameters, the bubble profile parameters, the initial bubble number, the simulation box length, and the number of Hubble lengths contained in the box at the end of the simulation.}
    \label{tab:para}
\end{table}

\section{Numerical Results}

\begin{figure}
    \centering
    \includegraphics[width=0.45\linewidth]{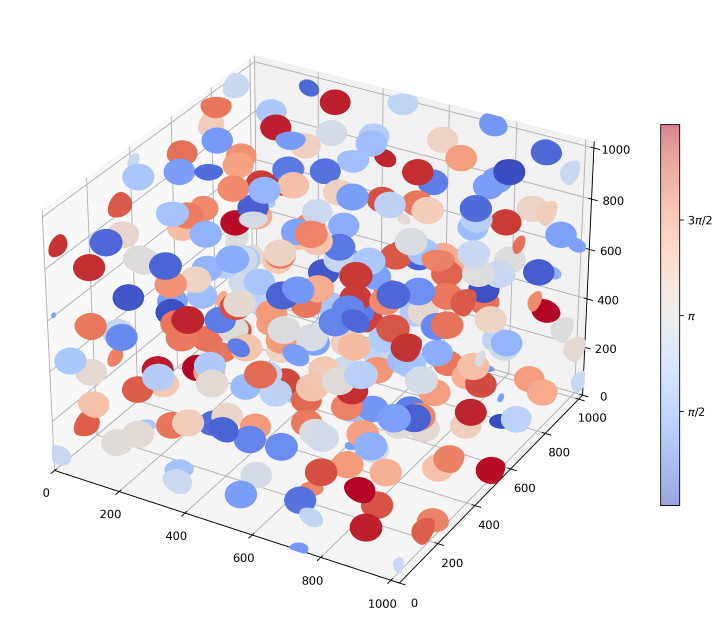}
    \includegraphics[width=0.45\linewidth]{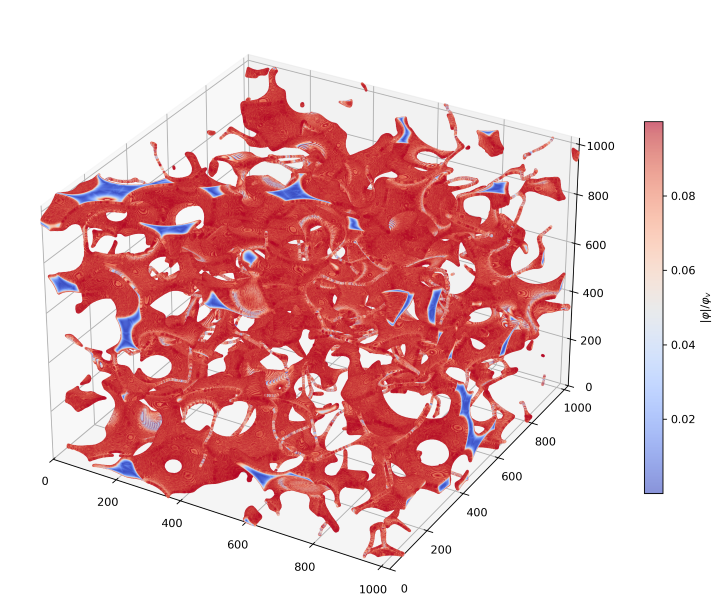} \\
    \includegraphics[width=0.42\linewidth]{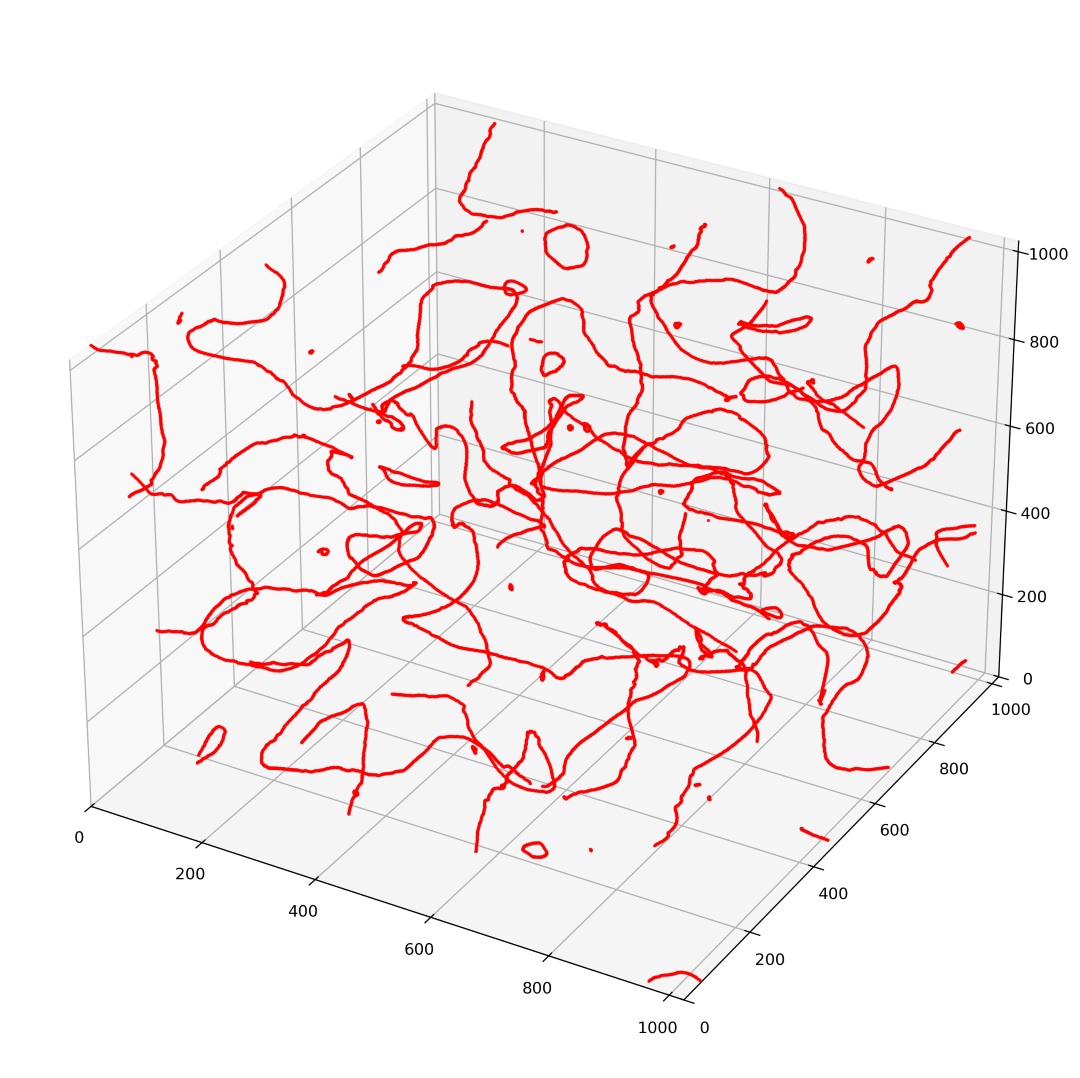}
    \includegraphics[width=0.42\linewidth]{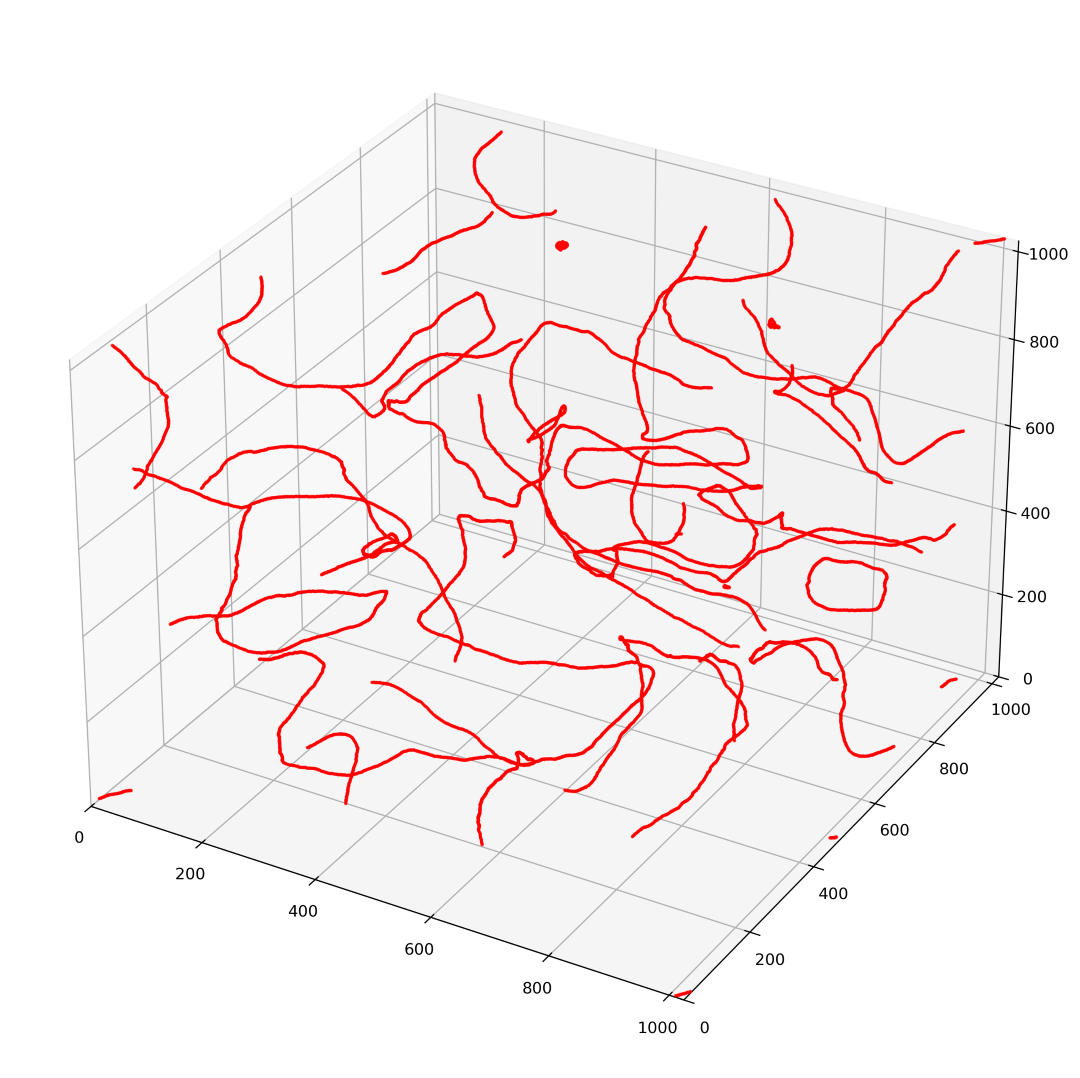}
    \caption{Top: snapshots of the phase field at $\tilde{\tau}=1.0$ (left), masked by $|\varphi| > 0.5 \varphi_v$, and the norm of complex field at $\tilde{\tau}=5.5$ (right), masked by $|\varphi| < 0.1 \varphi_v$. Bottom: snapshots of the string network at $\tilde{\tau}=8.5$ (left) and $11.5$ (right), using the phase identification method.}
    \label{fig:snapshot}
\end{figure}

Snapshots of the bubble and string evolution for setup 1 parameters are shown in Fig.~\ref{fig:snapshot}, which illustrates the formation of cosmic strings from bubble collisions. The upper panels display the phase distribution at initial rescaled times $\tilde{\tau}=1.0$ with a mask applied for $|\varphi| > 0.5\varphi_v$, and the residual true-vacuum regions. The lower panels show the identified string networks at $\tilde{\tau}=8.5$ and $11.5$, obtained via the phase identification method. One can observe that as the bubbles collide and merge, a string network appears in the remaining false-vacuum regions and persists.

\begin{figure}
    \centering
    \includegraphics[width=0.70\linewidth]{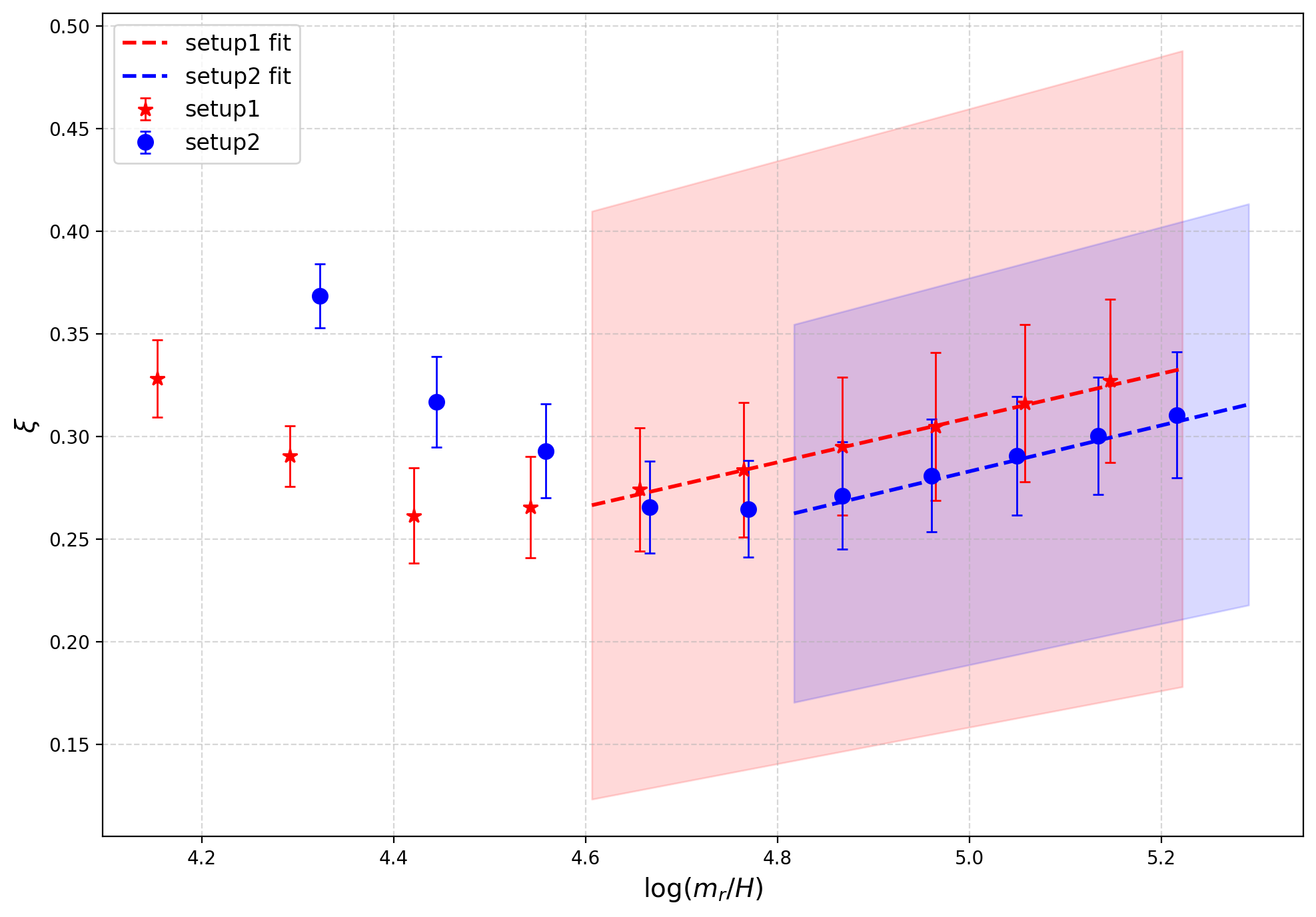}
    \caption{Evolution of the scaling parameter in the simulation. Symbols with error bars are the mean and standard deviation over five realizations. The dashed lines correspond to the fitting results with shaded bands from the fit-parameter uncertainties.}
    \label{fig:scaling_parameter}
\end{figure}

The evolution of the scaling parameter is presented in Fig.~\ref{fig:scaling_parameter}, following the method described in~\cite{Hiramatsu_2011,Jia:2024zeu}. The two first-order setups, characterized by different bubble numbers $N_b$, yield distinct initial string densities. It can be seen that, despite the effect of the initial density of global strings, the string network still enters the scaling regime. We fit the scaling parameter with $\xi = \alpha \log m_r/H + \beta$ and obtain $\alpha = 0.108 \pm 0.024$ and $\beta = -0.231 \pm 0.091$ for setup1, and $\alpha = 0.112 \pm 0.015$ and $\beta = -0.277 \pm 0.057$ for setup2, where the errors are the standard deviations among the five realizations. The value of $\alpha$ is smaller than the result $0.24(2), 0.254(2),0.21(2), ~\text{and}~ 0.205(7)/0.23(6)$
in the second order PT scenario~\cite{Gorghetto:2020qws,Buschmann_2022,Benabou:2024msj,Saikawa:2024bta}. 

\begin{figure}[!htp]
\centering
\includegraphics[width=0.80\linewidth]{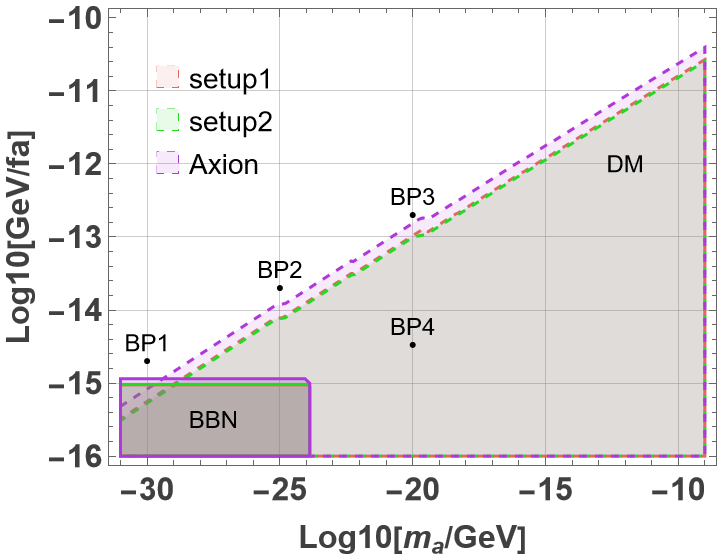}
\caption{Constraints on the ALPs parameter $f_a$ and $m_a$ from dark matter relic density and dark radiation. The axion case corresponds to the result in Ref.~\cite{Jia:2024zeu}. The shaded region is excluded.}
\label{fig:darkmatter}
\end{figure}

We calculate the axions emitted from the global string network using the method in Refs.~\cite{Hiramatsu_2011,Jia:2024zeu}.  
The mean inverse comoving momentum $k^{-1}(t)$ is obtained from the dimensionless parameter $\epsilon^{-1}$, and the scaling parameter is computed using the values of $\alpha$ and $\beta$. After the string stage, the number density scales as $n \propto a^{-3}$ until today without considering other effects, and these axions can be regarded as part of dark matter. Moreover, if the string ends after BBN, the energy density of free axions is constrained by the upper bound of dark radiation $\Delta N_{eff} < 0.46$. see Appendix.~\ref{app:axion} for details. The parameter space of axion-like particles (ALPs), constrained by dark matter relic density $\Omega h^2 <0.12$ and dark radiation, is shown in Fig.~\ref{fig:darkmatter}. Compared with the constraints on axion-like particles produced in the second-order PT~\cite{Jia:2024zeu}, the constraints on $f_a$ and $m_a$ from dark matter and BBN are slightly weaker.

\begin{figure}[!htp]
    \centering
    \includegraphics[width=0.8\linewidth]{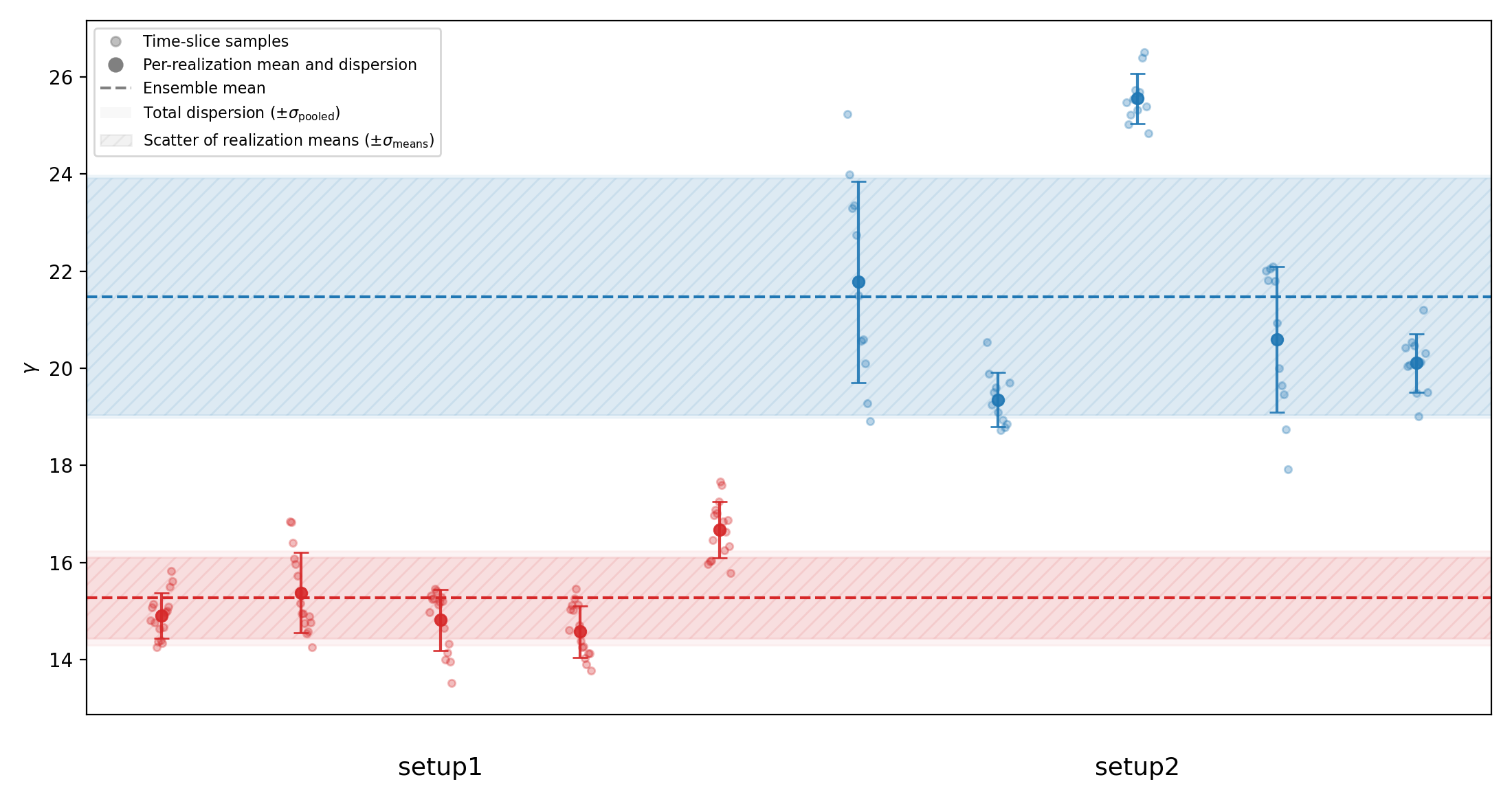}
    \caption{ The ratio $\gamma$ between the axion and GWs emission.}
    \label{fig:ratioaxiongw}
\end{figure}

\begin{figure}
\centering
\includegraphics[width=0.80\linewidth]{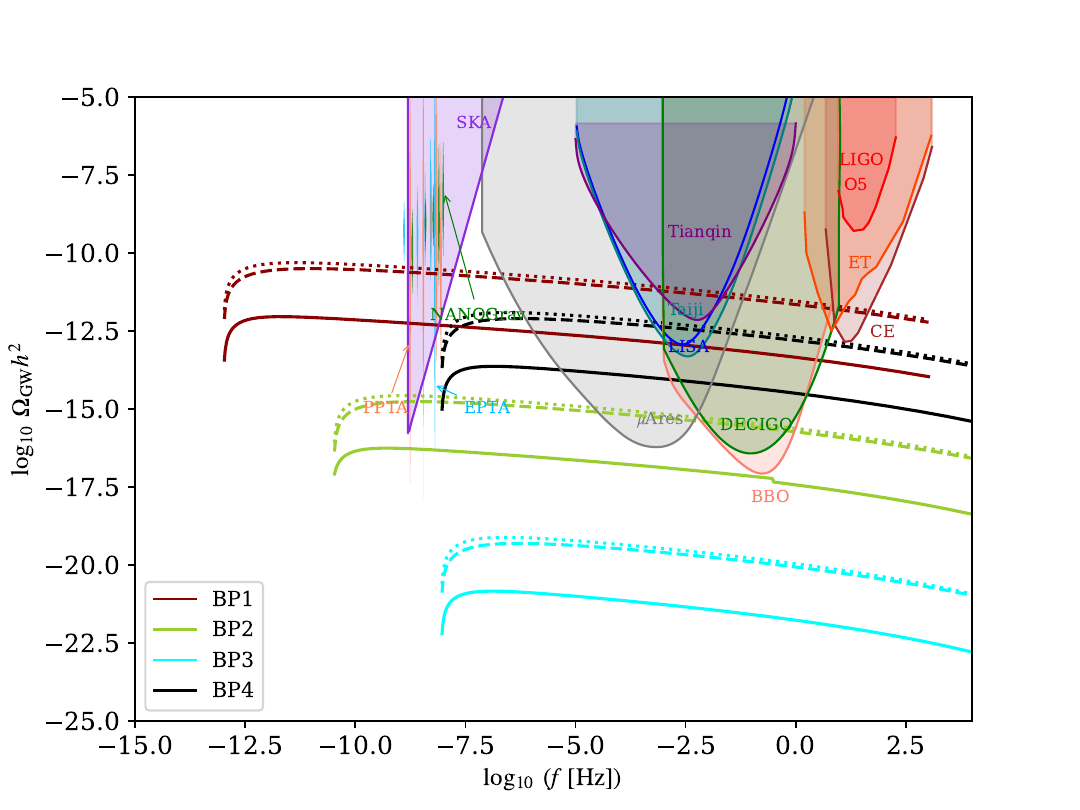}
\caption{The gravitational waves spectrum today for the values of $f_a$ and $m_a$ chosen in Fig.~\ref{fig:darkmatter}. The dashed and dotted lines correspond to the setup1 and setup2 inputs, respectively. The solid line corresponds to the result in Ref.~\cite{Jia:2024zeu}. }
\label{fig:gwtoday}
\end{figure}

We use the instantaneous emission spectrum $F_g(k/H,m_r/H)$ and the ratio $\gamma$ to obtain the GWs spectrum during the string stage. The GW emission rate is related to the axion emission rate by $\Gamma_{gw} \simeq \gamma \Gamma_a G \mu^2/f_a^2 $ with $\Gamma_a = a^{-4}\frac{\mathrm{d} (a^4 \rho_a)}{\mathrm{d}t}$ and $\Gamma_{gw} = a^{-4}\frac{\mathrm{d} (a^4 \rho_{gw})}{\mathrm{d}t}$. The forward difference is taken over a time interval $\tilde{\tau} = 1.0$ with initial times $\tilde{\tau} = 10.0,11.5$, and the result is shown in Fig.~\ref{fig:ratioaxiongw}. The value of $\gamma$ is $ 15.27 \pm 0.84, 21.48\pm 2.44$ for setups 1 and 2, respectively. See Appendix.~\ref{app:gws power spectrum} for detailed calculation of GWs spectrum and the instantaneous emission spectrum $F_g$.
After redshiftting, the present-day GWs spectrum is shown in Fig.~\ref{fig:gwtoday}, with GW detectors such as LISA~\cite{LISA:2017pwj,Baker:2019nia}, TianQin~\cite{TianQin:2015yph,Zhou:2023rop}, Taiji~\cite{Hu:2017mde,Ruan:2018tsw}, $\mu Ares$~\cite{Sesana:2019vho}, DECIGO~\cite{Seto:2001qf,Kawamura:2011zz,Yagi:2011wg,Isoyama:2018rjb}, BBO~\cite{Crowder:2005nr,Corbin:2005ny,Harry:2006fi}, SKA~\cite{Janssen:2014dka}, ET~\cite{Punturo:2010zz,Hild:2010id}, CE~\cite{LIGOScientific:2016wof} and LIGO~\cite{LIGOScientific:2014qfs,Thrane:2013oya,LIGOScientific:2019vic}. 
We find that the GWs spectra are two orders of magnitude higher than those of the axion string with the second-order PTs. The reason is as follows: the emission of GWs is a small part of the string evolution, with $\gamma \sim O(0.1)$ obtained in the second-order PTs~\cite{Gorghetto:2021fsn,Jia:2024zeu}; however, the $\gamma$ factors for the two setups of the first-order PTs are two orders higher than that cases.

\section{Conclusion and discussion}

In this work, we have studied the formation and evolution of global cosmic strings produced by a cosmological first-order PT, together with the free axions and the GWs emitted by the resulting string network.
As the bubbles merge, the false vacuum is compressed into sheets that shrink into line-like defects, leaving behind a string network.
We find that first-order PTs yield a slightly smaller scaling parameter than the second-order scenarios. This leads to weaker constraints on the PQ scale ($f_a$) and the axion-like particle mass ($m_a$) from dark matter relic density and BBN. 
Furthermore, the GW-to-axion emission ratio is enhanced by about two orders of magnitude compared with the second-order PT. 
In contrast to the conventional expectation for global strings from second-order PTs~\cite{Gorghetto:2020qws,Gorghetto:2021fsn,Gorghetto_2018}, the enhanced GW channel offers an opportunity to probe the parameter space of the axion decay constant $f_a$ and mass $m_a$ for global strings using near-future GW detectors such as PTA, LISA, and $\mu$Ares.

Future studies may explore phase equilibration during the evolution and how it can be described. Since thermal-plasma friction is not included in our setup, the bubble walls expand freely; this allows the dependence of the string formation efficiency on the bubble wall velocity to be investigated. Local strings from a first-order PT are also worth studying to determine their formation probability and GW emission efficiency. Moreover, the axion density from strings may affect axion miniclusters and axion stars~\cite{Ellis:2022grh,Eggemeier:2024fzs}; we leave these aspects for future study.

\section{Acknowledgements}
This work is supported by the National Natural Science Foundation of China (NSFC) under Grants Nos. 12322505, 12547101, and 12675130. L.B. also acknowledges Chongqing Natural Science Foundation under Grant
No. CSTB2024NSCQ-JQX0022 and Chongqing Talents: Exceptional Young Talents Project No. cstc2024ycjh-bgzxm0020. H.-K. G. is
supported by the startup fund provided by the University of Chinese Academy of Sciences
and by the NSFC under Grants No. 12547104 and No. 12475109.

\appendix

\begin{figure}[!htp]
    \centering
    \includegraphics[width=0.8\linewidth]{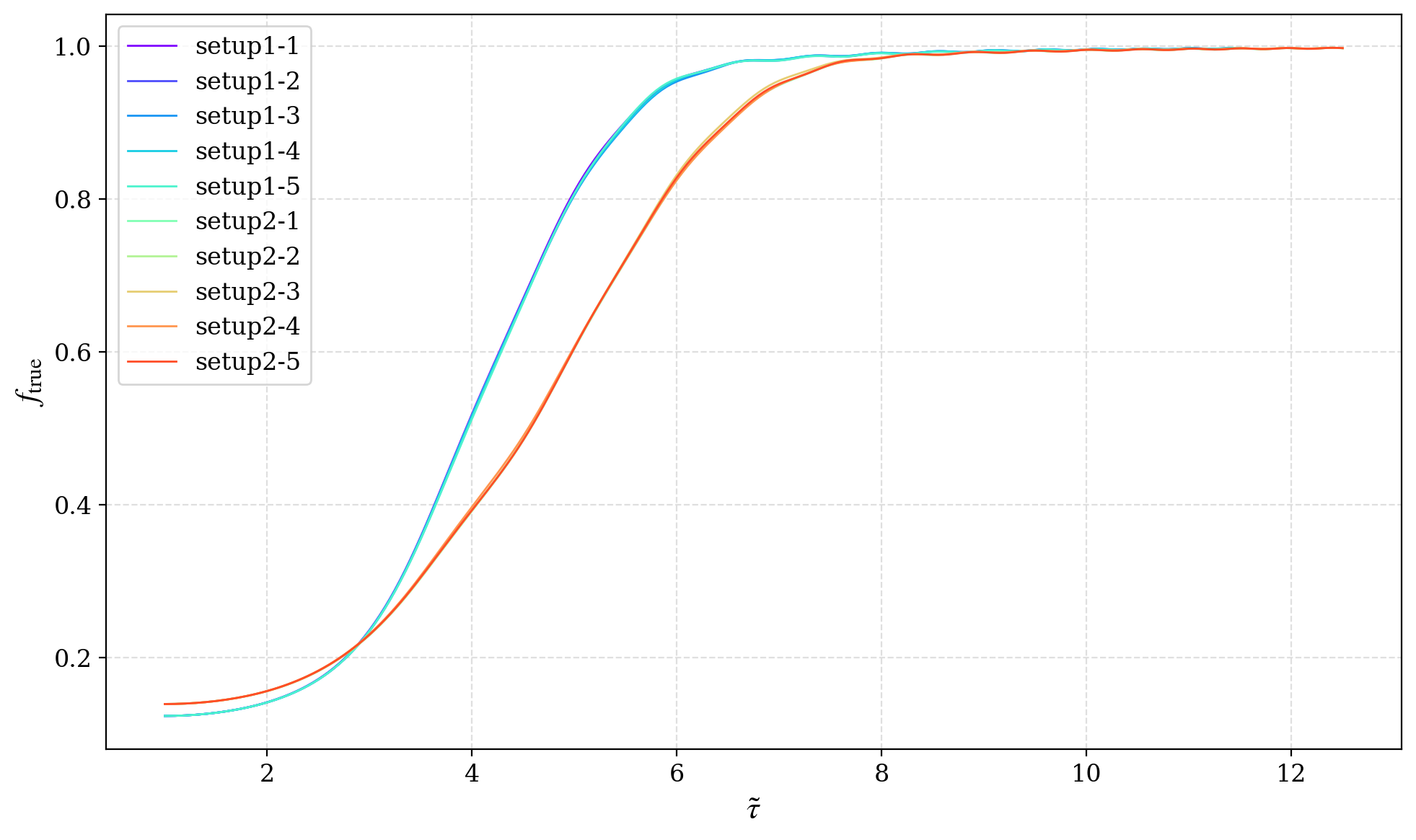}
    \includegraphics[width=0.8\linewidth]{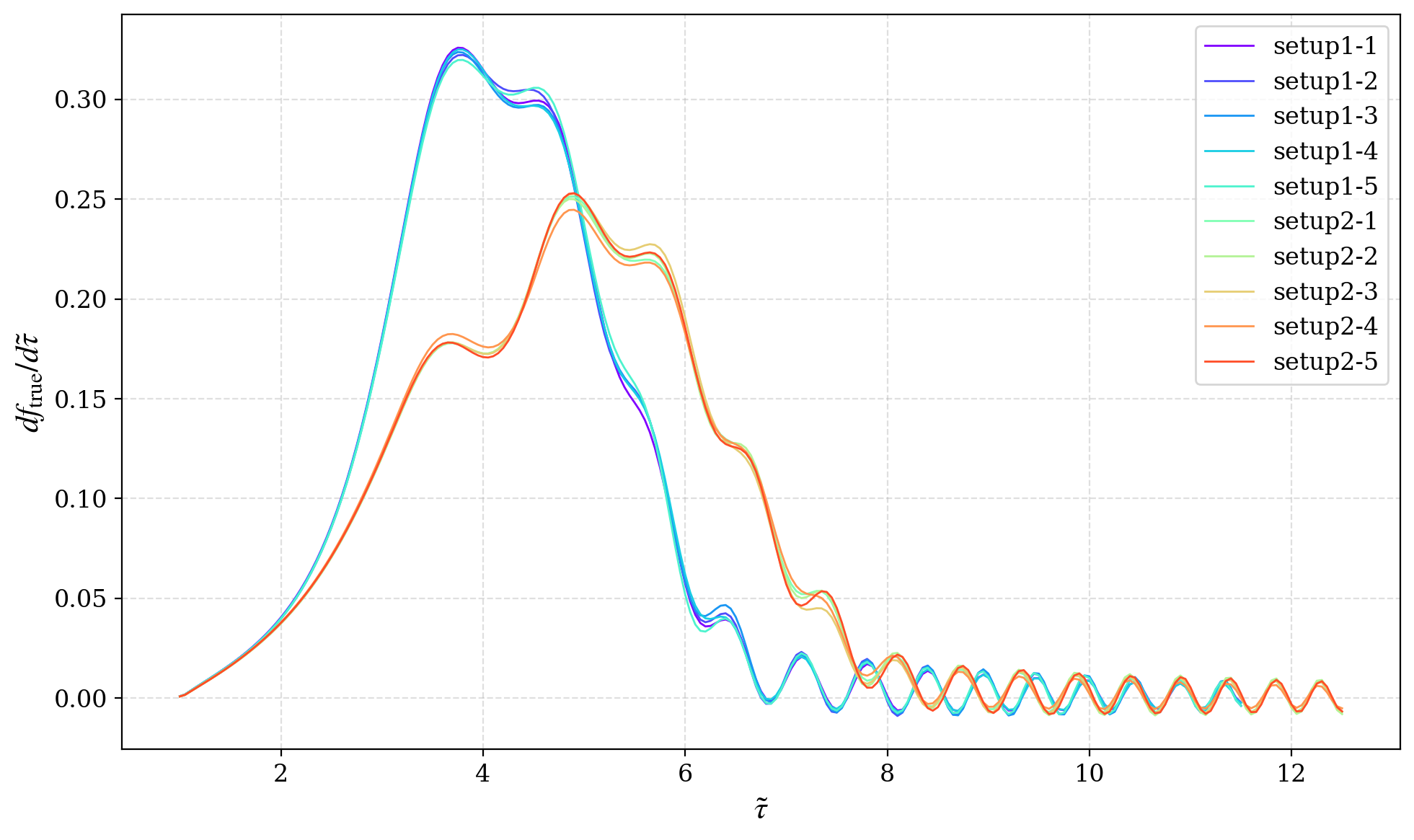}
    \includegraphics[width=0.8\linewidth]{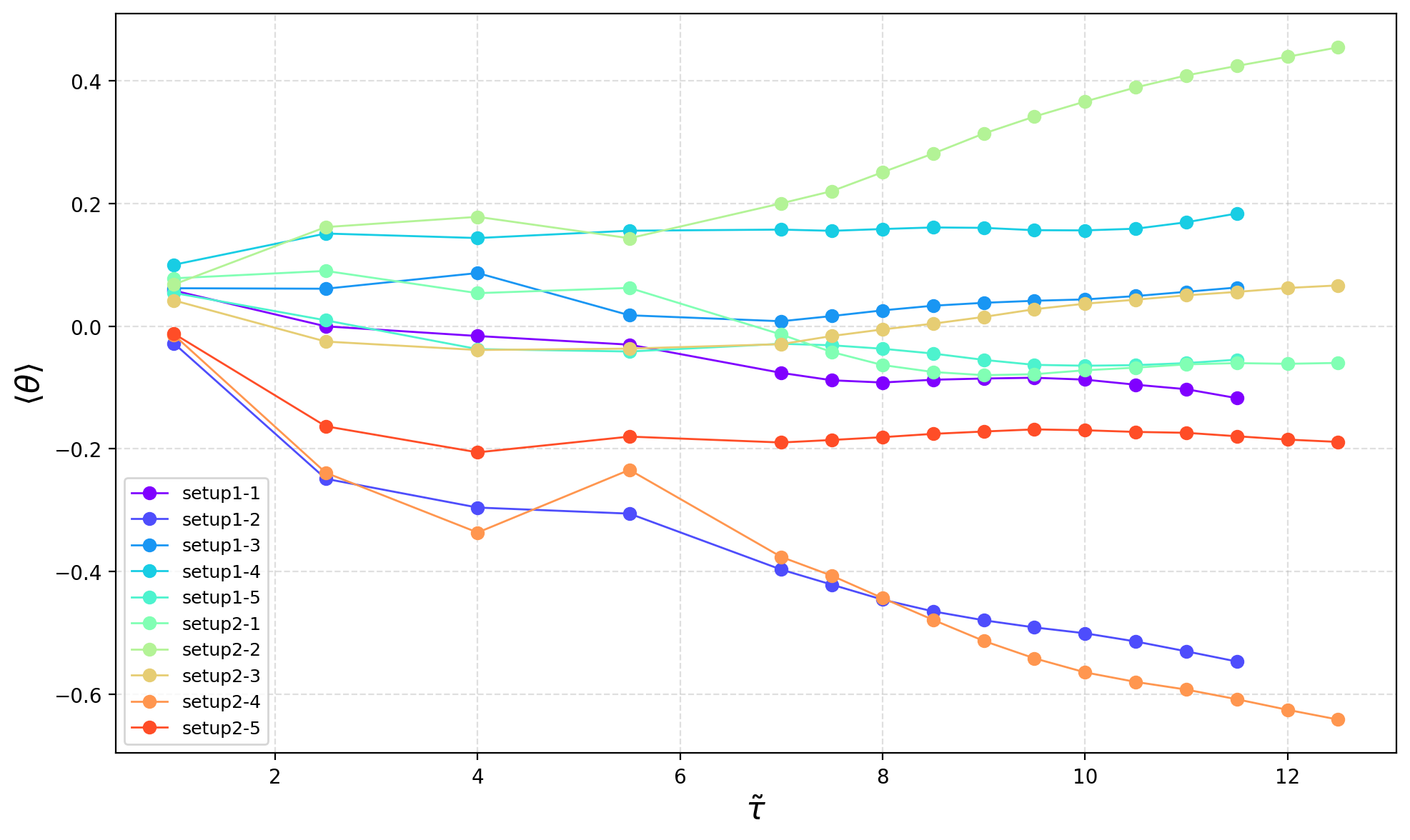}
    \caption{Evolution of the true-vacuum fractions (top), their time derivatives (middle), and the average phases (bottom).}
    \label{fig:field_value}
\end{figure}

\section{Vacuum and phase evolutions}
\label{sec:vp}
We show the evolutions of the true-vacuum fraction, $f_{true} = \langle|\varphi|\rangle / \varphi_v$, its time derivative, and the average phase with respect to the comoving rescaled time in Fig.~\ref{fig:field_value}. Setup 1 has a larger initial bubble number and completes the first-order PT earlier. And, for both parameter setups,, the average phases for different realizations have different departures from the $zero$, although the true-vacuum fractions and their time derivatives are all almost the same.

\section{Free axion spectra}
\label{app:axion}
\begin{figure}[!htp]
    \centering
    \includegraphics[width=0.8\linewidth]{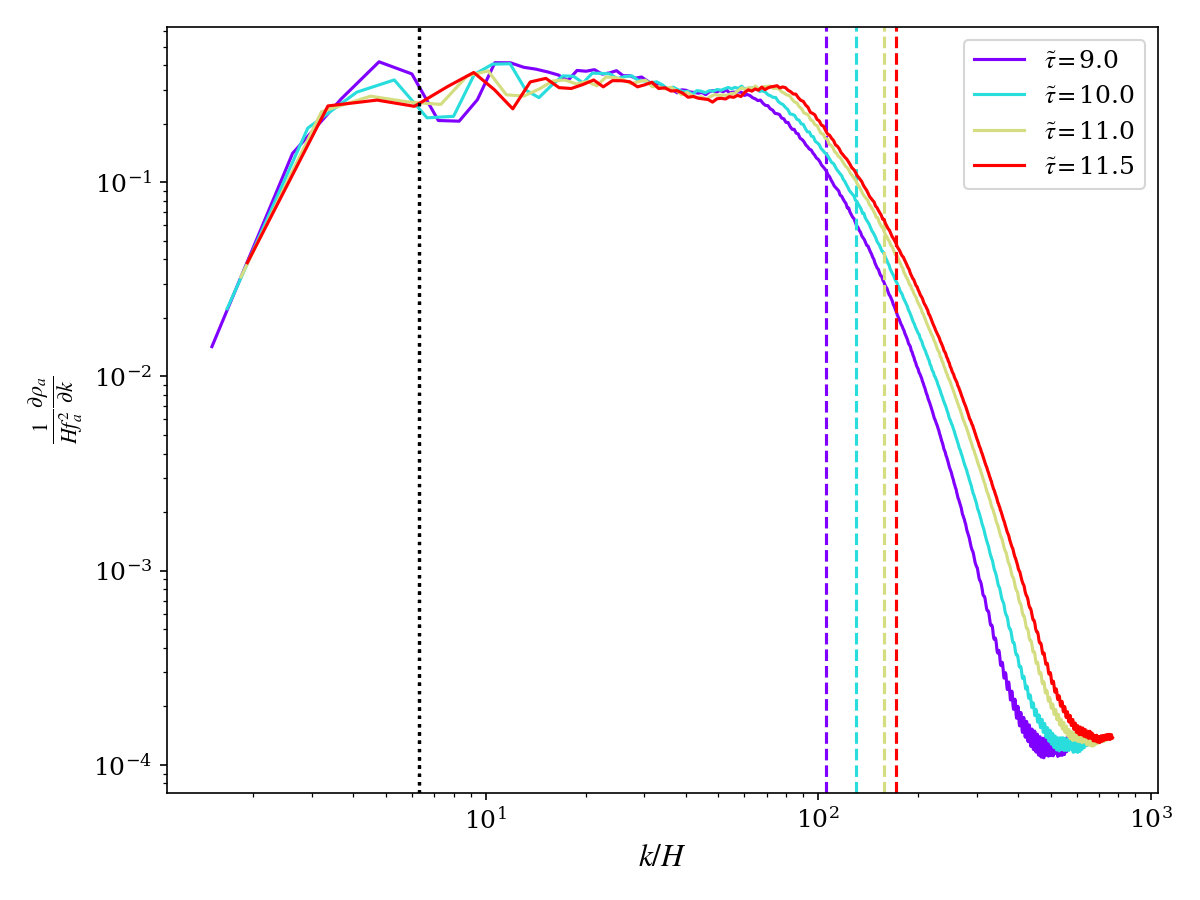}
    \includegraphics[width=0.8\linewidth]{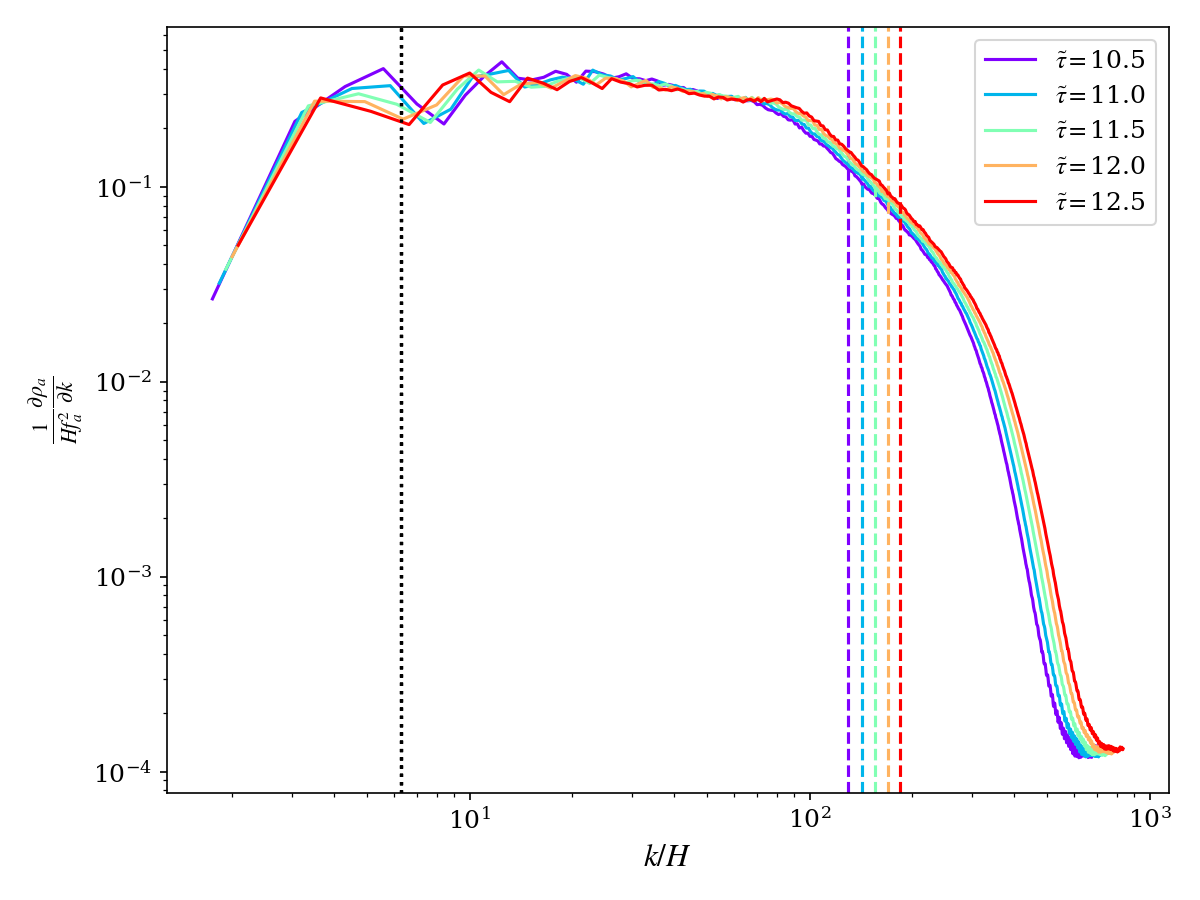}
    \caption{Evolution of the axion spectra for setups 1 and 2 (from top to bottom). The results are averaged over five realizations.}
    \label{fig:axion_spec}
\end{figure}

For the free axion emission spectrum, the energy density of free axions can be decomposed as $\rho_a \approx  \langle \dot{a}_s^2 \rangle = \int \frac{\partial \rho_a}{\partial |\mathbf{k}|} \, \mathrm{d}|\mathbf{k}|$, where the masked axion field is defined as $\dot{a}_s(x,t) = \dot{a}(x,t)|\varphi| / |\varphi_v|$. This mask equals 0 near the string core and approaches 1 far from the string, thereby excluding the contribution from string cores. The spectrum is presented in Fig.~\ref{fig:axion_spec}, where one can see that it changes little after the string enters the scaling regime.

Under the assumption that the mean physical reciprocal momentum of free axion is proportional to the Hubble scale, the dimensionless parameters can be extracted from simulation by $\epsilon^{-1} = a(t) k^{-1}(t)/(t/2\pi)$. The comoving mean reciprocal momentum of axion, $ k^{-1}(t)$, is computed as \cite{Hiramatsu_2011}
\begin{equation}
    k^{-1}(t) = \frac{\int \frac{\mathrm{d}k}{2\pi^2}\frac{1}{k}\Delta P(k,t)}{\int \frac{\mathrm{d}k}{2\pi^2}\Delta P(k,t)}\;,
\end{equation} 
where the differential spectrum is defined by $\Delta P(k,t) = a(t_2)^4P(k,t_2)-a(t_1)^4P(k,t_1)$, and $P(k,t)$ is the free axion power spectrum in the comoving momentum space. The inverse slope parameter $\epsilon^{-1} = 0.494 \pm 0.023, 0.430\pm0.073$ is extracted from the data at begin time $\tilde{\tau}=9.0,10.5$ and the end time $\tilde{\tau} =11.5,12.5$. 

In the evolution of the string network, the energy conversion of global strings can be described by
\begin{align}
    &\frac{ \mathrm{d}\rho_{s}(t)}{ \mathrm{d}t} = - 2 H(t) \rho_{s}(t) -  \left[\frac{\mathrm{d} \rho_{s}(t)}{\mathrm{d}t}\right]_{a} - \left[\frac{\mathrm{d} \rho_{s}(t)}{\mathrm{d}t}\right]_{g}\;, \\
    &\frac{ \mathrm{d}\rho_{a}(t)}{ \mathrm{d}t} = - 4 H(t) \rho_{a}(t) + \left[\frac{\mathrm{d} \rho_{s}(t)}{\mathrm{d}t}\right]_{a} \;, \\ \label{eq:enqx}
    &\frac{ \mathrm{d}\rho_{gw}(t)}{ \mathrm{d}t} = - 4 H(t) \rho_{gw}(t) + \left[\frac{\mathrm{d} \rho_{s}(t)}{\mathrm{d}t}\right]_{g} \;. 
\end{align}
where the energy of the string network is converted into free axions and gravitational waves. The string energy can be described by $ \rho_s = \pi f_a^2 \ln(t/(\sqrt{\xi}d_s)) \xi / t^2$, and the GW emission term can be neglected when $f_a \ll m_{pl} $ in the axion energy calculation. The axion number density is calculated by  
\begin{equation}
    N(t)=\int^{t_w}_{t_*} \mathrm{d}t k^{-1}(t) \frac{\mathrm{d}E(t)}{\mathrm{d}t} \;,
\label{eq:axionnumtw}
\end{equation}
where $N(t) \equiv a^3n_{a}(t)$ and $E(t) \equiv a^4\rho_{a}(t)$ are the comoving number and energy density of axion. The integration is performed from the time when $T = f_a$ to the time when $m_a = 3H$.
The effective number of neutrinos in BBN is calculated as $\Delta N_{eff} = (8/7)(11/4)^{4/3} \rho_a/\rho_\gamma$ at temperature of $1 \text{MeV}$ for free axion energy density.

For the QCD case, the axion mass squared is given by $m_a^2=\alpha (T/\Lambda)^{-n} \Lambda^4/f_a^2$, where we use the parameters $\alpha =1.68*10^{-7}, \Lambda = 400 ~\text{MeV}, n = 6.68$ from Ref.~\cite{Wantz_2010}. The present QCD axion energy density from strings is calculated by $\Omega_{a,string} = n_{a,0} m_{a,0} / \rho_{crit,0}$, where the zero-temperature axion mass is $m_a^2(T=0)=1.46\times10^{-3}\Lambda^4/f_a^2$, and $\rho_{crit,0}$ is the critical density today. The total axion energy density also includes axions from the misalignment mechanism~\cite{PhysRevD.91.065014}. The dark matter relic density constrains the QCD axion decay constant to $f_a <1.2\times10^{10}, 1.3\times10^{10} \text{GeV}$ for setups 1-2, respectively. 

\begin{figure}
    \centering
    \includegraphics[width=0.70\linewidth]{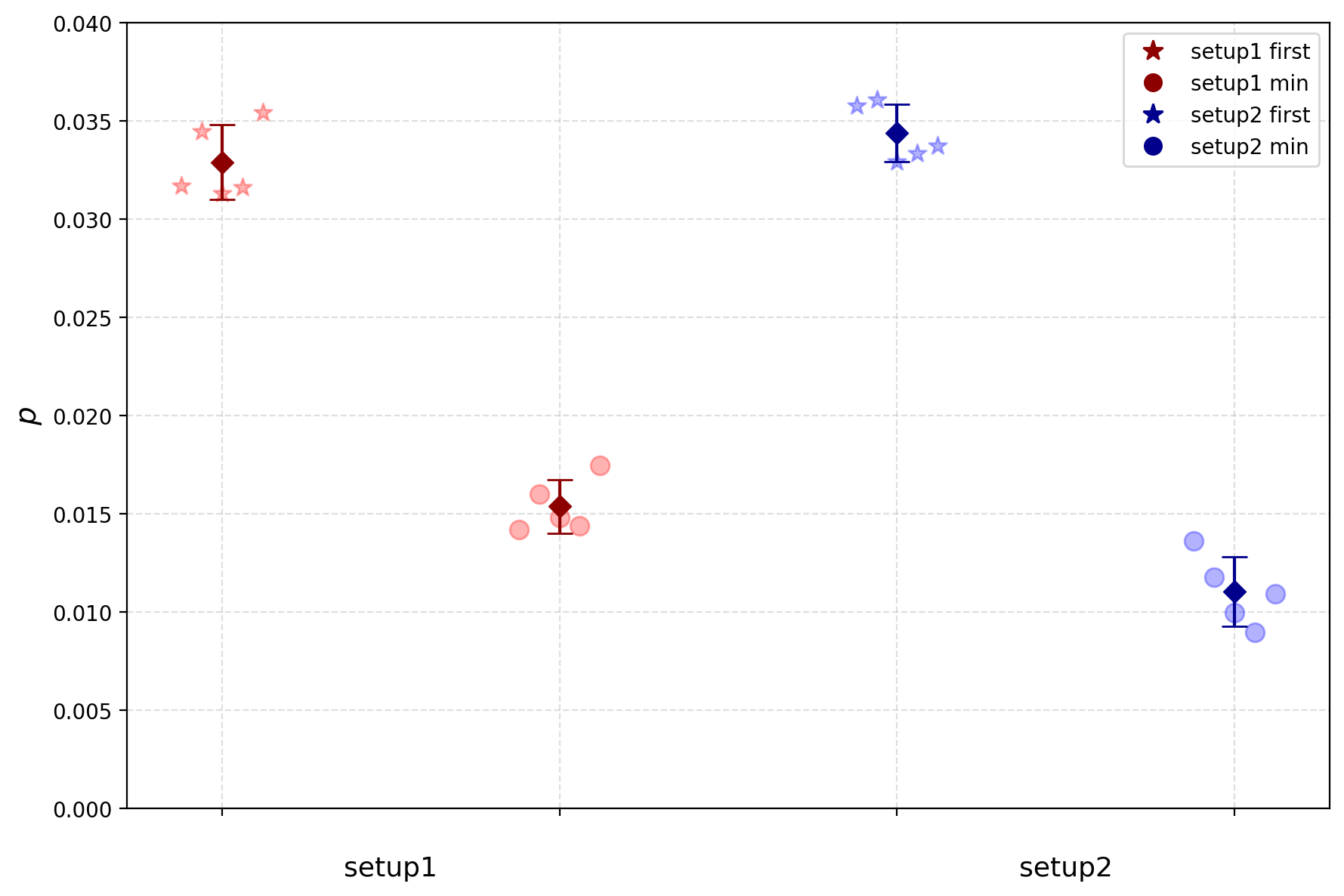}
    \caption{The ratio of strings generated from bubble collisions. The star denotes data at the beginning of the scaling parameter, and the solid circle denotes data at the time corresponding to the minimum scaling parameter.}
    \label{fig:bubblestringratio}
\end{figure}

We also calculate the generation ratio $ p = \ell_{s}/ R_{*}^{-2}$, where $\ell_{s} = \xi /t^2$ and $R_* = (N_b/V)^{-1/3}$, shown in Fig.~\ref{fig:bubblestringratio}, using data taken at the onset of the scaling regime and at the time of the minimum scaling parameter. This ratio is much less than the theoretically predicted value of  $0.25$ for the collisions of three bubbles ~\cite{Kibble:1995aa}. 

\section{Gravitational wave spectra evaluation}
\label{app:gws power spectrum}
GWs correspond to the transverse-traceless(TT) component of the metric perturbation $h_{ij}$, defined by $h_{ij} = g_{ij} - \delta_{ij}$. 
The equation of motion for GWs is
\begin{equation}
    \ddot{h}_{ij} + 3H\dot{h}_{ij} - \frac{1}{a^2} \nabla^2 h_{ij} = \frac{2}{m_{pl}^2a^2}\Pi^{TT}_{ij}	
    \label{eq:gws} \;,
\end{equation}
where an overdot denotes differentiation with respect to cosmic time, $m_{pl}$ is the reduced Planck mass, and $\Pi^{TT}_{ij}$ is the transverse-traceless part of the anisotropic tensor $\Pi_{ij}$. 
The anisotropic tensor is given by $\Pi_{ij} = T_{ij} - pg_{ij}$, where the energy-momentum tensor is evaluated as $T_{ij} = 2Re\left[\partial_i \varphi \; \partial_j \varphi^*\right]$\cite{Figueroa:2021yhd}, while the background pressure term does not contribute to $\Pi^{TT}_{ij}$. In Fourier space, the transverse-traceless anisotropic tensor is obtained via $\Pi_{ij}^{TT} = \Lambda_{ij,kl} \Pi_{kl}$, with the projection operator defined as
\begin{align}
    \Lambda_{ij,kl}(\hat{\textbf{k}}) &= P_{ik}({\hat{\textbf{k}}}) P_{jl}({\hat{\textbf{k}}})-\frac{1}{2}P_{ij}({\hat{\textbf{k}}})P_{kl}({\hat{\textbf{k}}}) \;,\nonumber \\
    P_{ij}({\hat{\textbf{k}}}) &= \delta_{ij} - \hat{k}_i\hat{k}_j \;,
\end{align}
where $\hat{k}_i = k_i / k$ is the unit vector along $\textbf{k}$. 
\begin{figure}[!htp]
    \centering
    \includegraphics[width=0.8\linewidth]{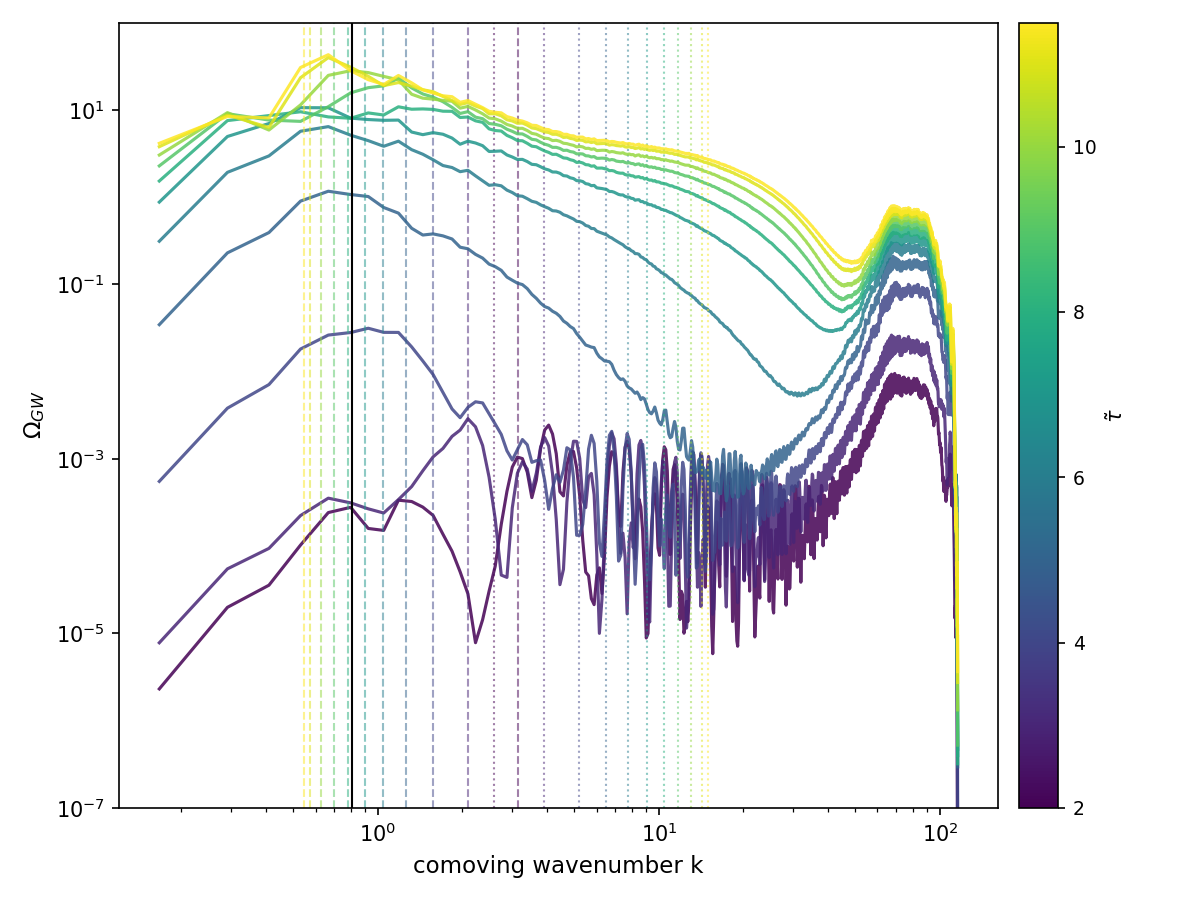} 
    \includegraphics[width=0.8\linewidth]{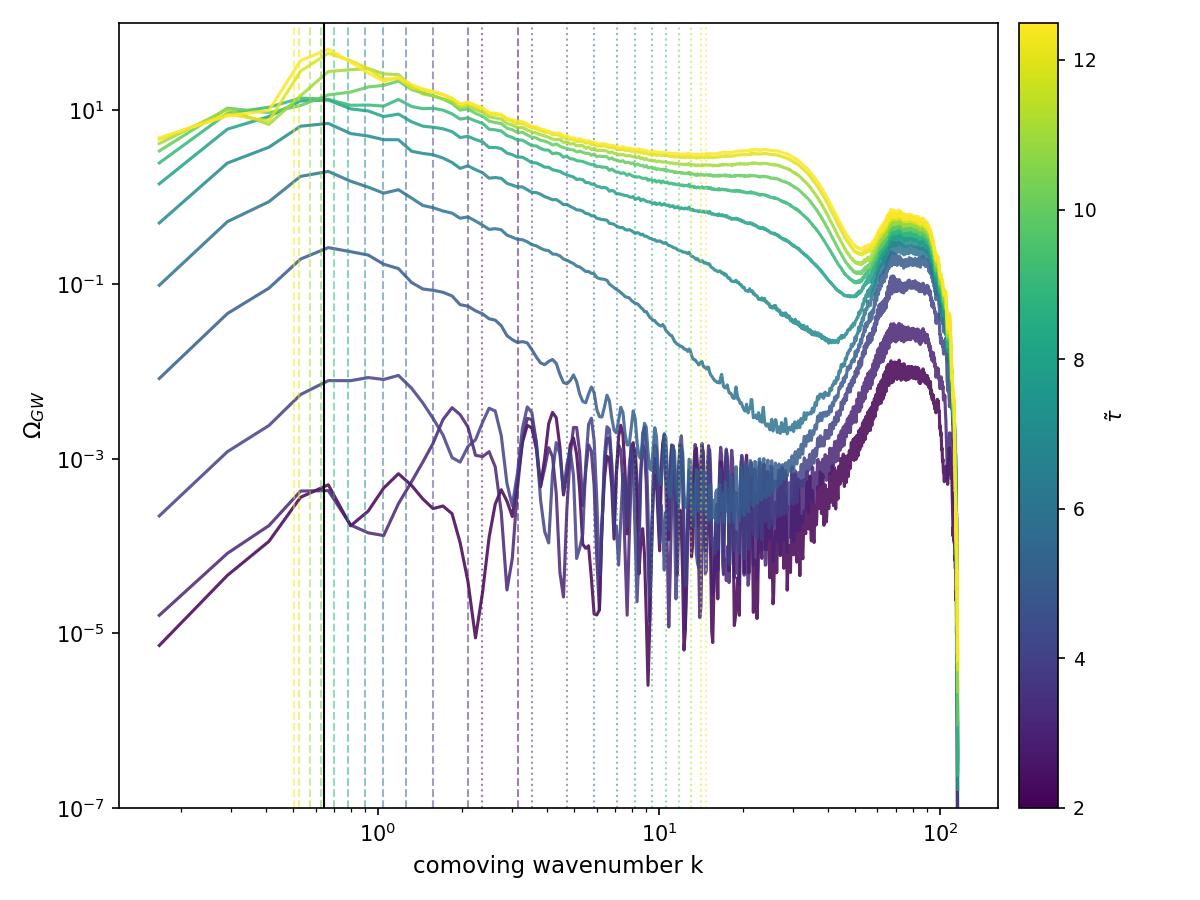} 
    \caption{Gravitational-wave spectra from the simulations for setups 1 and 2 (from top to bottom), averaged over five realizations. The same color solid line corresponds to the same rescaled conformal time $\tau$. The dashed line and dotted line correspond to $2 \pi H$ and the mass of the radial mode $m_r$. The black solid line correspond to $2 \pi R^{-1}_*$}
    \label{fig:gws}
\end{figure}
For convenience in numerically computing GWs in lattice simulation, the auxiliary tensor $u_{ij}$ is introduced and evolved, with the metric perturbation recovered as $h_{ij} = \Lambda_{ij,kl} u_{ij}$. The equation of motion of $u_{ij}$ is
\begin{align}
    \ddot{u}_{ij} + 3H\dot{u}_{ij} - \frac{1}{a^2} \nabla^2 u_{ij} = \frac{2}{m_{pl}^2a^2}\Pi_{ij} \;.
\end{align}
The gravitational wave energy density is defined as 
\begin{align}
    \rho_{gw}(t) = \frac{1}{32 \pi G}\left<\hat{h}_{ij}(\textbf{x},t) \hat{h}_{ij}(\textbf{x},t) \right> \;, 
\end{align}
where $\left<...\right>$ denotes spatial averaging. 
The gravitational wave spectrum generated by the bubble collisions and subsequent string dynamics is shown in Fig.~\ref{fig:gws}. 
One can see that bubble expansion and collisions cause the GW spectrum to increase significantly before the strings enter the scaling regime; after entering the scaling regime, the GW spectrum increases slowly due to string evolution. 
\begin{figure}[!htp]
    \centering
    \includegraphics[width=0.8\linewidth]{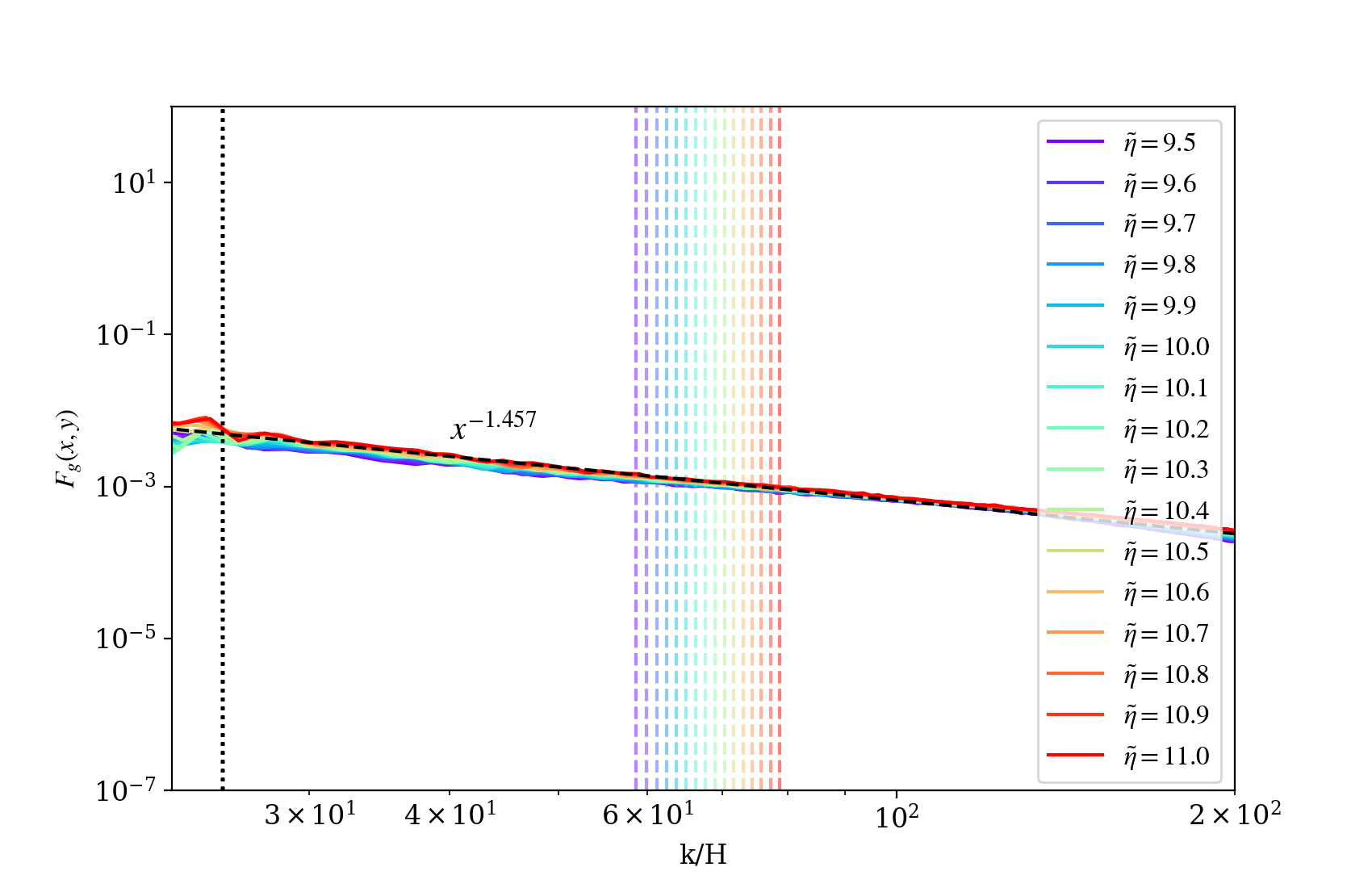}
    \includegraphics[width=0.8\linewidth]{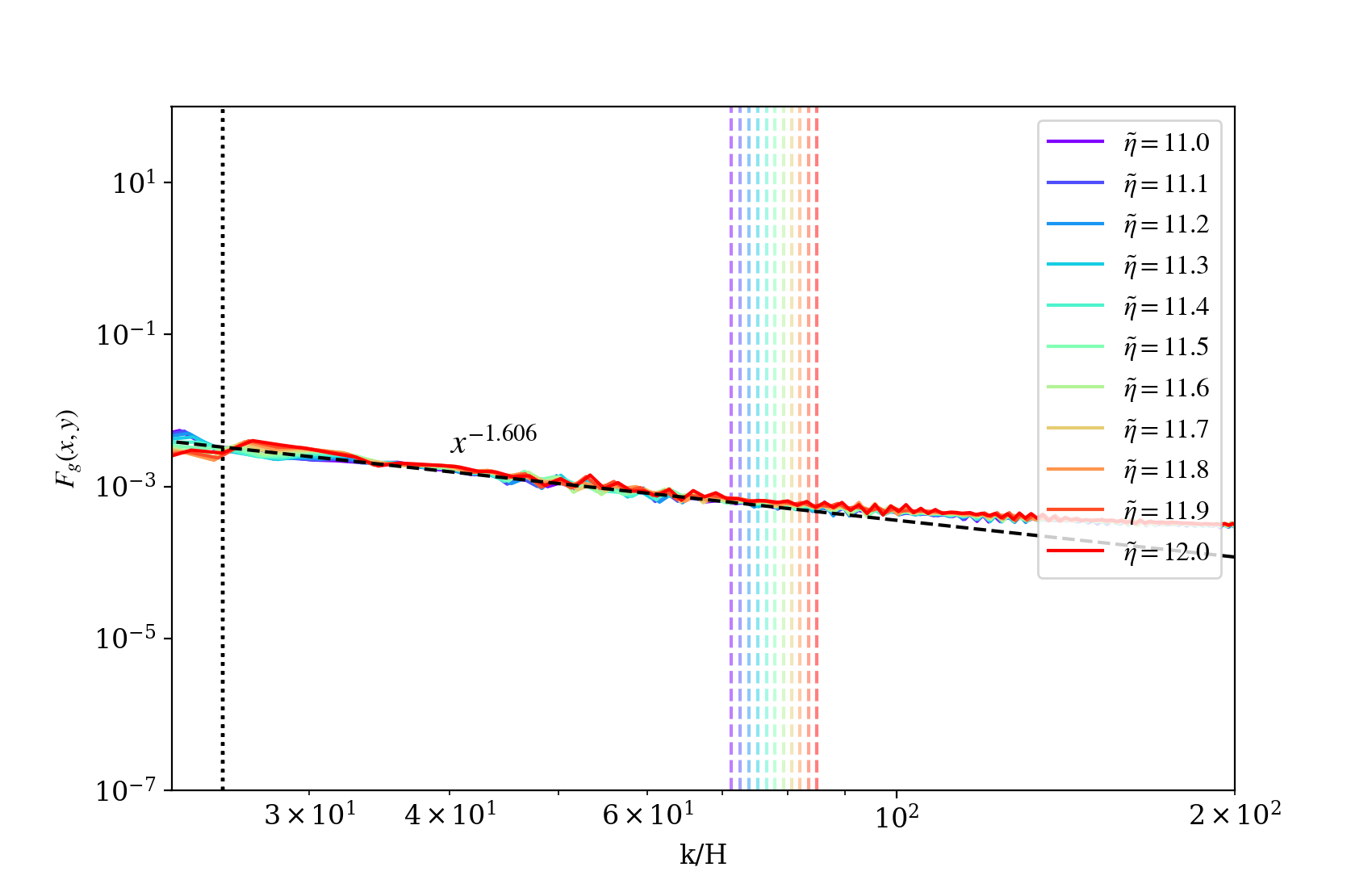}  
    \caption{Evolution of the instantaneous emission spectrum $F(x,y)$ for setups 1 and 2 (from top to bottom), averaged over five realizations. The dotted black line corresponds to $8 \pi H$, and the dashed lines correspond to $m_r/2$. The black dashed line corresponds to the fitting result.}
    \label{fig:gws_emi}
\end{figure}

The methods in \cite{Gorghetto:2021fsn,Jia:2024zeu} are used to calculate the GWs during the string evolution epoch. The instantaneous emission spectrum $F_g(x,y)$ is calculated by:
\begin{align}
    F_g(x,y)=\frac{H}{\Gamma_{gw}}\frac{1}{a^3}\frac{\partial}{\partial t} \left(a^3\frac{\partial \rho_{gw}}{\partial k} \right) \;,
\end{align} 
where the variables are defined as $x=k/H$ and $y=m_r/H$. The spectral index is obtained by fitting $F_g(x,y) \propto x^{-q}$, yielding  $q = 1.457 \pm 0.037$ and $ 1.606 \pm 0.066$ for setups 1 and 2, respectively, in the range $[8 \pi H, m_r /2 ]$, using the same differencing method and data as in the $\gamma$ calculation. The results are presented in Fig.~\ref{fig:gws_emi}.

Together with the axion energy, the GWs spectrum at the end time of the global string is calculated from the emission spectrum $F_g(x,y)$ and the fitting parameters $\gamma$ and $q$. The GWs spectrum is then redshifted as $\Omega_{GW,0} = \Omega_{\text{rad},0}(g_0/g_e)^{\frac{1}{3}} \Omega_{GW,e}$, and the physical frequency scales as $ k \propto a^{-1}$ with the redshift factor $a_0^4/a_e^4=g_e^{1/3}\rho_{\text{rad},e}/(g_0^{1/3}\rho_{\text{rad},0})$ \cite{Jia:2024zeu}.

\bibliography{ref}

\begin{thebibliography}{81}%
\makeatletter
\providecommand \@ifxundefined [1]{%
 \@ifx{#1\undefined}
}%
\providecommand \@ifnum [1]{%
 \ifnum #1\expandafter \@firstoftwo
 \else \expandafter \@secondoftwo
 \fi
}%
\providecommand \@ifx [1]{%
 \ifx #1\expandafter \@firstoftwo
 \else \expandafter \@secondoftwo
 \fi
}%
\providecommand \natexlab [1]{#1}%
\providecommand \enquote  [1]{``#1''}%
\providecommand \bibnamefont  [1]{#1}%
\providecommand \bibfnamefont [1]{#1}%
\providecommand \citenamefont [1]{#1}%
\providecommand \href@noop [0]{\@secondoftwo}%
\providecommand \href [0]{\begingroup \@sanitize@url \@href}%
\providecommand \@href[1]{\@@startlink{#1}\@@href}%
\providecommand \@@href[1]{\endgroup#1\@@endlink}%
\providecommand \@sanitize@url [0]{\catcode `\\12\catcode `\$12\catcode
  `\&12\catcode `\#12\catcode `\^12\catcode `\_12\catcode `\%12\relax}%
\providecommand \@@startlink[1]{}%
\providecommand \@@endlink[0]{}%
\providecommand \url  [0]{\begingroup\@sanitize@url \@url }%
\providecommand \@url [1]{\endgroup\@href {#1}{\urlprefix }}%
\providecommand \urlprefix  [0]{URL }%
\providecommand \Eprint [0]{\href }%
\providecommand \doibase [0]{http://dx.doi.org/}%
\providecommand \selectlanguage [0]{\@gobble}%
\providecommand \bibinfo  [0]{\@secondoftwo}%
\providecommand \bibfield  [0]{\@secondoftwo}%
\providecommand \translation [1]{[#1]}%
\providecommand \BibitemOpen [0]{}%
\providecommand \bibitemStop [0]{}%
\providecommand \bibitemNoStop [0]{.\EOS\space}%
\providecommand \EOS [0]{\spacefactor3000\relax}%
\providecommand \BibitemShut  [1]{\csname bibitem#1\endcsname}%
\let\auto@bib@innerbib\@empty
\bibitem [{\citenamefont {Agazie}\ \emph {et~al.}(2023)\citenamefont {Agazie}
  \emph {et~al.}}]{NANOGrav:2023gor}%
  \BibitemOpen
  \bibfield  {author} {\bibinfo {author} {\bibfnamefont {G.}~\bibnamefont
  {Agazie}} \emph {et~al.} (\bibinfo {collaboration} {NANOGrav}),\ }\href
  {\doibase 10.3847/2041-8213/acdac6} {\bibfield  {journal} {\bibinfo
  {journal} {Astrophys. J. Lett.}\ }\textbf {\bibinfo {volume} {951}},\
  \bibinfo {pages} {L8} (\bibinfo {year} {2023})},\ \Eprint
  {http://arxiv.org/abs/2306.16213} {arXiv:2306.16213 [astro-ph.HE]}
  \BibitemShut {NoStop}%
\bibitem [{\citenamefont {Antoniadis}\ \emph {et~al.}(2023)\citenamefont
  {Antoniadis} \emph {et~al.}}]{EPTA:2023fyk}%
  \BibitemOpen
  \bibfield  {author} {\bibinfo {author} {\bibfnamefont {J.}~\bibnamefont
  {Antoniadis}} \emph {et~al.} (\bibinfo {collaboration} {EPTA, InPTA:}),\
  }\href {\doibase 10.1051/0004-6361/202346844} {\bibfield  {journal} {\bibinfo
   {journal} {Astron. Astrophys.}\ }\textbf {\bibinfo {volume} {678}},\
  \bibinfo {pages} {A50} (\bibinfo {year} {2023})},\ \Eprint
  {http://arxiv.org/abs/2306.16214} {arXiv:2306.16214 [astro-ph.HE]}
  \BibitemShut {NoStop}%
\bibitem [{\citenamefont {Reardon}\ \emph {et~al.}(2023)\citenamefont {Reardon}
  \emph {et~al.}}]{Reardon:2023gzh}%
  \BibitemOpen
  \bibfield  {author} {\bibinfo {author} {\bibfnamefont {D.~J.}\ \bibnamefont
  {Reardon}} \emph {et~al.},\ }\href {\doibase 10.3847/2041-8213/acdd02}
  {\bibfield  {journal} {\bibinfo  {journal} {Astrophys. J. Lett.}\ }\textbf
  {\bibinfo {volume} {951}},\ \bibinfo {pages} {L6} (\bibinfo {year} {2023})},\
  \Eprint {http://arxiv.org/abs/2306.16215} {arXiv:2306.16215 [astro-ph.HE]}
  \BibitemShut {NoStop}%
\bibitem [{\citenamefont {Xu}\ \emph {et~al.}(2023)\citenamefont {Xu} \emph
  {et~al.}}]{Xu:2023wog}%
  \BibitemOpen
  \bibfield  {author} {\bibinfo {author} {\bibfnamefont {H.}~\bibnamefont {Xu}}
  \emph {et~al.},\ }\href {\doibase 10.1088/1674-4527/acdfa5} {\bibfield
  {journal} {\bibinfo  {journal} {Res. Astron. Astrophys.}\ }\textbf {\bibinfo
  {volume} {23}},\ \bibinfo {pages} {075024} (\bibinfo {year} {2023})},\
  \Eprint {http://arxiv.org/abs/2306.16216} {arXiv:2306.16216 [astro-ph.HE]}
  \BibitemShut {NoStop}%
\bibitem [{\citenamefont {Kibble}(1976)}]{Kibble:1976sj}%
  \BibitemOpen
  \bibfield  {author} {\bibinfo {author} {\bibfnamefont {T.~W.~B.}\
  \bibnamefont {Kibble}},\ }\href {\doibase 10.1088/0305-4470/9/8/029}
  {\bibfield  {journal} {\bibinfo  {journal} {J. Phys. A}\ }\textbf {\bibinfo
  {volume} {9}},\ \bibinfo {pages} {1387} (\bibinfo {year} {1976})}\BibitemShut
  {NoStop}%
\bibitem [{\citenamefont {Vilenkin}\ and\ \citenamefont
  {Everett}(1982)}]{Vilenkin:1982ks}%
  \BibitemOpen
  \bibfield  {author} {\bibinfo {author} {\bibfnamefont {A.}~\bibnamefont
  {Vilenkin}}\ and\ \bibinfo {author} {\bibfnamefont {A.~E.}\ \bibnamefont
  {Everett}},\ }\href {\doibase 10.1103/PhysRevLett.48.1867} {\bibfield
  {journal} {\bibinfo  {journal} {Phys. Rev. Lett.}\ }\textbf {\bibinfo
  {volume} {48}},\ \bibinfo {pages} {1867} (\bibinfo {year}
  {1982})}\BibitemShut {NoStop}%
\bibitem [{\citenamefont {Kitajima}\ and\ \citenamefont
  {Nakayama}(2023)}]{Kitajima:2023vre}%
  \BibitemOpen
  \bibfield  {author} {\bibinfo {author} {\bibfnamefont {N.}~\bibnamefont
  {Kitajima}}\ and\ \bibinfo {author} {\bibfnamefont {K.}~\bibnamefont
  {Nakayama}},\ }\href {\doibase 10.1016/j.physletb.2023.138213} {\bibfield
  {journal} {\bibinfo  {journal} {Phys. Lett. B}\ }\textbf {\bibinfo {volume}
  {846}},\ \bibinfo {pages} {138213} (\bibinfo {year} {2023})},\ \Eprint
  {http://arxiv.org/abs/2306.17390} {arXiv:2306.17390 [hep-ph]} \BibitemShut
  {NoStop}%
\bibitem [{\citenamefont {Bian}\ \emph {et~al.}(2024)\citenamefont {Bian},
  \citenamefont {Ge}, \citenamefont {Shu}, \citenamefont {Wang}, \citenamefont
  {Yang},\ and\ \citenamefont {Zong}}]{Bian:2023dnv}%
  \BibitemOpen
  \bibfield  {author} {\bibinfo {author} {\bibfnamefont {L.}~\bibnamefont
  {Bian}}, \bibinfo {author} {\bibfnamefont {S.}~\bibnamefont {Ge}}, \bibinfo
  {author} {\bibfnamefont {J.}~\bibnamefont {Shu}}, \bibinfo {author}
  {\bibfnamefont {B.}~\bibnamefont {Wang}}, \bibinfo {author} {\bibfnamefont
  {X.-Y.}\ \bibnamefont {Yang}}, \ and\ \bibinfo {author} {\bibfnamefont
  {J.}~\bibnamefont {Zong}},\ }\href {\doibase 10.1103/PhysRevD.109.L101301}
  {\bibfield  {journal} {\bibinfo  {journal} {Phys. Rev. D}\ }\textbf {\bibinfo
  {volume} {109}},\ \bibinfo {pages} {L101301} (\bibinfo {year} {2024})},\
  \Eprint {http://arxiv.org/abs/2307.02376} {arXiv:2307.02376 [astro-ph.HE]}
  \BibitemShut {NoStop}%
\bibitem [{\citenamefont {Basilakos}\ \emph {et~al.}(2024)\citenamefont
  {Basilakos}, \citenamefont {Nanopoulos}, \citenamefont {Papanikolaou},
  \citenamefont {Saridakis},\ and\ \citenamefont
  {Tzerefos}}]{Basilakos:2023xof}%
  \BibitemOpen
  \bibfield  {author} {\bibinfo {author} {\bibfnamefont {S.}~\bibnamefont
  {Basilakos}}, \bibinfo {author} {\bibfnamefont {D.~V.}\ \bibnamefont
  {Nanopoulos}}, \bibinfo {author} {\bibfnamefont {T.}~\bibnamefont
  {Papanikolaou}}, \bibinfo {author} {\bibfnamefont {E.~N.}\ \bibnamefont
  {Saridakis}}, \ and\ \bibinfo {author} {\bibfnamefont {C.}~\bibnamefont
  {Tzerefos}},\ }\href {\doibase 10.1016/j.physletb.2024.138507} {\bibfield
  {journal} {\bibinfo  {journal} {Phys. Lett. B}\ }\textbf {\bibinfo {volume}
  {850}},\ \bibinfo {pages} {138507} (\bibinfo {year} {2024})},\ \Eprint
  {http://arxiv.org/abs/2307.08601} {arXiv:2307.08601 [hep-th]} \BibitemShut
  {NoStop}%
\bibitem [{\citenamefont {Wang}\ \emph {et~al.}(2023)\citenamefont {Wang},
  \citenamefont {Lei}, \citenamefont {Jiao}, \citenamefont {Feng},\ and\
  \citenamefont {Fan}}]{Wang:2023len}%
  \BibitemOpen
  \bibfield  {author} {\bibinfo {author} {\bibfnamefont {Z.}~\bibnamefont
  {Wang}}, \bibinfo {author} {\bibfnamefont {L.}~\bibnamefont {Lei}}, \bibinfo
  {author} {\bibfnamefont {H.}~\bibnamefont {Jiao}}, \bibinfo {author}
  {\bibfnamefont {L.}~\bibnamefont {Feng}}, \ and\ \bibinfo {author}
  {\bibfnamefont {Y.-Z.}\ \bibnamefont {Fan}},\ }\href {\doibase
  10.1007/s11433-023-2262-0} {\bibfield  {journal} {\bibinfo  {journal} {Sci.
  China Phys. Mech. Astron.}\ }\textbf {\bibinfo {volume} {66}},\ \bibinfo
  {pages} {120403} (\bibinfo {year} {2023})},\ \Eprint
  {http://arxiv.org/abs/2306.17150} {arXiv:2306.17150 [astro-ph.HE]}
  \BibitemShut {NoStop}%
\bibitem [{\citenamefont {Ellis}\ \emph {et~al.}(2023)\citenamefont {Ellis},
  \citenamefont {Lewicki}, \citenamefont {Lin},\ and\ \citenamefont
  {Vaskonen}}]{Ellis:2023tsl}%
  \BibitemOpen
  \bibfield  {author} {\bibinfo {author} {\bibfnamefont {J.}~\bibnamefont
  {Ellis}}, \bibinfo {author} {\bibfnamefont {M.}~\bibnamefont {Lewicki}},
  \bibinfo {author} {\bibfnamefont {C.}~\bibnamefont {Lin}}, \ and\ \bibinfo
  {author} {\bibfnamefont {V.}~\bibnamefont {Vaskonen}},\ }\href {\doibase
  10.1103/PhysRevD.108.103511} {\bibfield  {journal} {\bibinfo  {journal}
  {Phys. Rev. D}\ }\textbf {\bibinfo {volume} {108}},\ \bibinfo {pages}
  {103511} (\bibinfo {year} {2023})},\ \Eprint
  {http://arxiv.org/abs/2306.17147} {arXiv:2306.17147 [astro-ph.CO]}
  \BibitemShut {NoStop}%
\bibitem [{\citenamefont {Buchmuller}\ \emph {et~al.}(2023)\citenamefont
  {Buchmuller}, \citenamefont {Domcke},\ and\ \citenamefont
  {Schmitz}}]{Buchmuller:2023aus}%
  \BibitemOpen
  \bibfield  {author} {\bibinfo {author} {\bibfnamefont {W.}~\bibnamefont
  {Buchmuller}}, \bibinfo {author} {\bibfnamefont {V.}~\bibnamefont {Domcke}},
  \ and\ \bibinfo {author} {\bibfnamefont {K.}~\bibnamefont {Schmitz}},\ }\href
  {\doibase 10.1088/1475-7516/2023/11/020} {\bibfield  {journal} {\bibinfo
  {journal} {JCAP}\ }\textbf {\bibinfo {volume} {11}},\ \bibinfo {pages} {020}
  (\bibinfo {year} {2023})},\ \Eprint {http://arxiv.org/abs/2307.04691}
  {arXiv:2307.04691 [hep-ph]} \BibitemShut {NoStop}%
\bibitem [{\citenamefont {Figueroa}\ \emph {et~al.}(2024)\citenamefont
  {Figueroa}, \citenamefont {Pieroni}, \citenamefont {Ricciardone},\ and\
  \citenamefont {Simakachorn}}]{Figueroa:2023zhu}%
  \BibitemOpen
  \bibfield  {author} {\bibinfo {author} {\bibfnamefont {D.~G.}\ \bibnamefont
  {Figueroa}}, \bibinfo {author} {\bibfnamefont {M.}~\bibnamefont {Pieroni}},
  \bibinfo {author} {\bibfnamefont {A.}~\bibnamefont {Ricciardone}}, \ and\
  \bibinfo {author} {\bibfnamefont {P.}~\bibnamefont {Simakachorn}},\ }\href
  {\doibase 10.1103/PhysRevLett.132.171002} {\bibfield  {journal} {\bibinfo
  {journal} {Phys. Rev. Lett.}\ }\textbf {\bibinfo {volume} {132}},\ \bibinfo
  {pages} {171002} (\bibinfo {year} {2024})},\ \Eprint
  {http://arxiv.org/abs/2307.02399} {arXiv:2307.02399 [astro-ph.CO]}
  \BibitemShut {NoStop}%
\bibitem [{\citenamefont {Afzal}\ \emph {et~al.}(2023)\citenamefont {Afzal}
  \emph {et~al.}}]{NANOGrav:2023hvm}%
  \BibitemOpen
  \bibfield  {author} {\bibinfo {author} {\bibfnamefont {A.}~\bibnamefont
  {Afzal}} \emph {et~al.} (\bibinfo {collaboration} {NANOGrav}),\ }\href
  {\doibase 10.3847/2041-8213/acdc91} {\bibfield  {journal} {\bibinfo
  {journal} {Astrophys. J. Lett.}\ }\textbf {\bibinfo {volume} {951}},\
  \bibinfo {pages} {L11} (\bibinfo {year} {2023})},\ \bibinfo {note} {[Erratum:
  Astrophys.J.Lett. 971, L27 (2024), Erratum: Astrophys.J. 971, L27 (2024)]},\
  \Eprint {http://arxiv.org/abs/2306.16219} {arXiv:2306.16219 [astro-ph.HE]}
  \BibitemShut {NoStop}%
\bibitem [{\citenamefont {Antoniadis}\ \emph {et~al.}(2024)\citenamefont
  {Antoniadis} \emph {et~al.}}]{EPTA:2023xxk}%
  \BibitemOpen
  \bibfield  {author} {\bibinfo {author} {\bibfnamefont {J.}~\bibnamefont
  {Antoniadis}} \emph {et~al.} (\bibinfo {collaboration} {EPTA, InPTA}),\
  }\href {\doibase 10.1051/0004-6361/202347433} {\bibfield  {journal} {\bibinfo
   {journal} {Astron. Astrophys.}\ }\textbf {\bibinfo {volume} {685}},\
  \bibinfo {pages} {A94} (\bibinfo {year} {2024})},\ \Eprint
  {http://arxiv.org/abs/2306.16227} {arXiv:2306.16227 [astro-ph.CO]}
  \BibitemShut {NoStop}%
\bibitem [{\citenamefont {Cheng}\ and\ \citenamefont
  {Visinelli}(2024)}]{Cheng:2024axj}%
  \BibitemOpen
  \bibfield  {author} {\bibinfo {author} {\bibfnamefont {H.}~\bibnamefont
  {Cheng}}\ and\ \bibinfo {author} {\bibfnamefont {L.}~\bibnamefont
  {Visinelli}},\ }\href {\doibase 10.1016/j.dark.2024.101667} {\bibfield
  {journal} {\bibinfo  {journal} {Phys. Dark Univ.}\ }\textbf {\bibinfo
  {volume} {46}},\ \bibinfo {pages} {101667} (\bibinfo {year} {2024})},\
  \Eprint {http://arxiv.org/abs/2408.16334} {arXiv:2408.16334 [hep-ph]}
  \BibitemShut {NoStop}%
\bibitem [{\citenamefont {Bian}\ \emph {et~al.}(2021)\citenamefont {Bian},
  \citenamefont {Cai}, \citenamefont {Liu}, \citenamefont {Yang},\ and\
  \citenamefont {Zhou}}]{Bian:2020urb}%
  \BibitemOpen
  \bibfield  {author} {\bibinfo {author} {\bibfnamefont {L.}~\bibnamefont
  {Bian}}, \bibinfo {author} {\bibfnamefont {R.-G.}\ \bibnamefont {Cai}},
  \bibinfo {author} {\bibfnamefont {J.}~\bibnamefont {Liu}}, \bibinfo {author}
  {\bibfnamefont {X.-Y.}\ \bibnamefont {Yang}}, \ and\ \bibinfo {author}
  {\bibfnamefont {R.}~\bibnamefont {Zhou}},\ }\href {\doibase
  10.1103/PhysRevD.103.L081301} {\bibfield  {journal} {\bibinfo  {journal}
  {Phys. Rev. D}\ }\textbf {\bibinfo {volume} {103}},\ \bibinfo {pages}
  {L081301} (\bibinfo {year} {2021})},\ \Eprint
  {http://arxiv.org/abs/2009.13893} {arXiv:2009.13893 [astro-ph.CO]}
  \BibitemShut {NoStop}%
\bibitem [{\citenamefont {Peccei}\ and\ \citenamefont
  {Quinn}(1977{\natexlab{a}})}]{Peccei:1977hh}%
  \BibitemOpen
  \bibfield  {author} {\bibinfo {author} {\bibfnamefont {R.~D.}\ \bibnamefont
  {Peccei}}\ and\ \bibinfo {author} {\bibfnamefont {H.~R.}\ \bibnamefont
  {Quinn}},\ }\href {\doibase 10.1103/PhysRevLett.38.1440} {\bibfield
  {journal} {\bibinfo  {journal} {Phys. Rev. Lett.}\ }\textbf {\bibinfo
  {volume} {38}},\ \bibinfo {pages} {1440} (\bibinfo {year}
  {1977}{\natexlab{a}})}\BibitemShut {NoStop}%
\bibitem [{\citenamefont {Peccei}\ and\ \citenamefont
  {Quinn}(1977{\natexlab{b}})}]{Peccei:1977ur}%
  \BibitemOpen
  \bibfield  {author} {\bibinfo {author} {\bibfnamefont {R.~D.}\ \bibnamefont
  {Peccei}}\ and\ \bibinfo {author} {\bibfnamefont {H.~R.}\ \bibnamefont
  {Quinn}},\ }\href {\doibase 10.1103/PhysRevD.16.1791} {\bibfield  {journal}
  {\bibinfo  {journal} {Phys. Rev. D}\ }\textbf {\bibinfo {volume} {16}},\
  \bibinfo {pages} {1791} (\bibinfo {year} {1977}{\natexlab{b}})}\BibitemShut
  {NoStop}%
\bibitem [{\citenamefont {Weinberg}(1978)}]{Weinberg:1977ma}%
  \BibitemOpen
  \bibfield  {author} {\bibinfo {author} {\bibfnamefont {S.}~\bibnamefont
  {Weinberg}},\ }\href {\doibase 10.1103/PhysRevLett.40.223} {\bibfield
  {journal} {\bibinfo  {journal} {Phys. Rev. Lett.}\ }\textbf {\bibinfo
  {volume} {40}},\ \bibinfo {pages} {223} (\bibinfo {year} {1978})}\BibitemShut
  {NoStop}%
\bibitem [{\citenamefont {Wilczek}(1978)}]{Wilczek:1977pj}%
  \BibitemOpen
  \bibfield  {author} {\bibinfo {author} {\bibfnamefont {F.}~\bibnamefont
  {Wilczek}},\ }\href {\doibase 10.1103/PhysRevLett.40.279} {\bibfield
  {journal} {\bibinfo  {journal} {Phys. Rev. Lett.}\ }\textbf {\bibinfo
  {volume} {40}},\ \bibinfo {pages} {279} (\bibinfo {year} {1978})}\BibitemShut
  {NoStop}%
\bibitem [{\citenamefont {Kim}(1979)}]{Kim:1979if}%
  \BibitemOpen
  \bibfield  {author} {\bibinfo {author} {\bibfnamefont {J.~E.}\ \bibnamefont
  {Kim}},\ }\href {\doibase 10.1103/PhysRevLett.43.103} {\bibfield  {journal}
  {\bibinfo  {journal} {Phys. Rev. Lett.}\ }\textbf {\bibinfo {volume} {43}},\
  \bibinfo {pages} {103} (\bibinfo {year} {1979})}\BibitemShut {NoStop}%
\bibitem [{\citenamefont {Shifman}\ \emph {et~al.}(1980)\citenamefont
  {Shifman}, \citenamefont {Vainshtein},\ and\ \citenamefont
  {Zakharov}}]{Shifman:1979if}%
  \BibitemOpen
  \bibfield  {author} {\bibinfo {author} {\bibfnamefont {M.~A.}\ \bibnamefont
  {Shifman}}, \bibinfo {author} {\bibfnamefont {A.~I.}\ \bibnamefont
  {Vainshtein}}, \ and\ \bibinfo {author} {\bibfnamefont {V.~I.}\ \bibnamefont
  {Zakharov}},\ }\href {\doibase 10.1016/0550-3213(80)90209-6} {\bibfield
  {journal} {\bibinfo  {journal} {Nucl. Phys. B}\ }\textbf {\bibinfo {volume}
  {166}},\ \bibinfo {pages} {493} (\bibinfo {year} {1980})}\BibitemShut
  {NoStop}%
\bibitem [{\citenamefont {Zhitnitsky}(1980)}]{Zhitnitsky:1980tq}%
  \BibitemOpen
  \bibfield  {author} {\bibinfo {author} {\bibfnamefont {A.~R.}\ \bibnamefont
  {Zhitnitsky}},\ }\href@noop {} {\bibfield  {journal} {\bibinfo  {journal}
  {Sov. J. Nucl. Phys.}\ }\textbf {\bibinfo {volume} {31}},\ \bibinfo {pages}
  {260} (\bibinfo {year} {1980})}\BibitemShut {NoStop}%
\bibitem [{\citenamefont {Dine}\ \emph {et~al.}(1981)\citenamefont {Dine},
  \citenamefont {Fischler},\ and\ \citenamefont {Srednicki}}]{Dine:1981rt}%
  \BibitemOpen
  \bibfield  {author} {\bibinfo {author} {\bibfnamefont {M.}~\bibnamefont
  {Dine}}, \bibinfo {author} {\bibfnamefont {W.}~\bibnamefont {Fischler}}, \
  and\ \bibinfo {author} {\bibfnamefont {M.}~\bibnamefont {Srednicki}},\ }\href
  {\doibase 10.1016/0370-2693(81)90590-6} {\bibfield  {journal} {\bibinfo
  {journal} {Phys. Lett. B}\ }\textbf {\bibinfo {volume} {104}},\ \bibinfo
  {pages} {199} (\bibinfo {year} {1981})}\BibitemShut {NoStop}%
\bibitem [{\citenamefont {Preskill}\ \emph {et~al.}(1983)\citenamefont
  {Preskill}, \citenamefont {Wise},\ and\ \citenamefont
  {Wilczek}}]{Preskill:1982cy}%
  \BibitemOpen
  \bibfield  {author} {\bibinfo {author} {\bibfnamefont {J.}~\bibnamefont
  {Preskill}}, \bibinfo {author} {\bibfnamefont {M.~B.}\ \bibnamefont {Wise}},
  \ and\ \bibinfo {author} {\bibfnamefont {F.}~\bibnamefont {Wilczek}},\ }\href
  {\doibase 10.1016/0370-2693(83)90637-8} {\bibfield  {journal} {\bibinfo
  {journal} {Phys. Lett. B}\ }\textbf {\bibinfo {volume} {120}},\ \bibinfo
  {pages} {127} (\bibinfo {year} {1983})}\BibitemShut {NoStop}%
\bibitem [{\citenamefont {Abbott}\ and\ \citenamefont
  {Sikivie}(1983)}]{Abbott:1982af}%
  \BibitemOpen
  \bibfield  {author} {\bibinfo {author} {\bibfnamefont {L.~F.}\ \bibnamefont
  {Abbott}}\ and\ \bibinfo {author} {\bibfnamefont {P.}~\bibnamefont
  {Sikivie}},\ }\href {\doibase 10.1016/0370-2693(83)90638-X} {\bibfield
  {journal} {\bibinfo  {journal} {Phys. Lett. B}\ }\textbf {\bibinfo {volume}
  {120}},\ \bibinfo {pages} {133} (\bibinfo {year} {1983})}\BibitemShut
  {NoStop}%
\bibitem [{\citenamefont {Dine}\ and\ \citenamefont
  {Fischler}(1983)}]{Dine:1982ah}%
  \BibitemOpen
  \bibfield  {author} {\bibinfo {author} {\bibfnamefont {M.}~\bibnamefont
  {Dine}}\ and\ \bibinfo {author} {\bibfnamefont {W.}~\bibnamefont
  {Fischler}},\ }\href {\doibase 10.1016/0370-2693(83)90639-1} {\bibfield
  {journal} {\bibinfo  {journal} {Phys. Lett. B}\ }\textbf {\bibinfo {volume}
  {120}},\ \bibinfo {pages} {137} (\bibinfo {year} {1983})}\BibitemShut
  {NoStop}%
\bibitem [{\citenamefont {Auclair}\ \emph {et~al.}(2020)\citenamefont
  {Auclair}, \citenamefont {Steer},\ and\ \citenamefont
  {Vachaspati}}]{Auclair:2019jip}%
  \BibitemOpen
  \bibfield  {author} {\bibinfo {author} {\bibfnamefont {P.}~\bibnamefont
  {Auclair}}, \bibinfo {author} {\bibfnamefont {D.~A.}\ \bibnamefont {Steer}},
  \ and\ \bibinfo {author} {\bibfnamefont {T.}~\bibnamefont {Vachaspati}},\
  }\href {\doibase 10.1103/PhysRevD.101.083511} {\bibfield  {journal} {\bibinfo
   {journal} {Phys. Rev. D}\ }\textbf {\bibinfo {volume} {101}},\ \bibinfo
  {pages} {083511} (\bibinfo {year} {2020})},\ \Eprint
  {http://arxiv.org/abs/1911.12066} {arXiv:1911.12066 [hep-ph]} \BibitemShut
  {NoStop}%
\bibitem [{\citenamefont {Matsunami}\ \emph {et~al.}(2019)\citenamefont
  {Matsunami}, \citenamefont {Pogosian}, \citenamefont {Saurabh},\ and\
  \citenamefont {Vachaspati}}]{Matsunami:2019fss}%
  \BibitemOpen
  \bibfield  {author} {\bibinfo {author} {\bibfnamefont {D.}~\bibnamefont
  {Matsunami}}, \bibinfo {author} {\bibfnamefont {L.}~\bibnamefont {Pogosian}},
  \bibinfo {author} {\bibfnamefont {A.}~\bibnamefont {Saurabh}}, \ and\
  \bibinfo {author} {\bibfnamefont {T.}~\bibnamefont {Vachaspati}},\ }\href
  {\doibase 10.1103/PhysRevLett.122.201301} {\bibfield  {journal} {\bibinfo
  {journal} {Phys. Rev. Lett.}\ }\textbf {\bibinfo {volume} {122}},\ \bibinfo
  {pages} {201301} (\bibinfo {year} {2019})},\ \Eprint
  {http://arxiv.org/abs/1903.05102} {arXiv:1903.05102 [hep-ph]} \BibitemShut
  {NoStop}%
\bibitem [{\citenamefont {Gorghetto}\ \emph {et~al.}(2018)\citenamefont
  {Gorghetto}, \citenamefont {Hardy},\ and\ \citenamefont
  {Villadoro}}]{Gorghetto_2018}%
  \BibitemOpen
  \bibfield  {author} {\bibinfo {author} {\bibfnamefont {M.}~\bibnamefont
  {Gorghetto}}, \bibinfo {author} {\bibfnamefont {E.}~\bibnamefont {Hardy}}, \
  and\ \bibinfo {author} {\bibfnamefont {G.}~\bibnamefont {Villadoro}},\ }\href
  {\doibase 10.1007/jhep07(2018)151} {\bibfield  {journal} {\bibinfo  {journal}
  {Journal of High Energy Physics}\ }\textbf {\bibinfo {volume} {2018}}
  (\bibinfo {year} {2018}),\ 10.1007/jhep07(2018)151}\BibitemShut {NoStop}%
\bibitem [{\citenamefont {Gorghetto}\ \emph
  {et~al.}(2021{\natexlab{a}})\citenamefont {Gorghetto}, \citenamefont
  {Hardy},\ and\ \citenamefont {Villadoro}}]{Gorghetto:2020qws}%
  \BibitemOpen
  \bibfield  {author} {\bibinfo {author} {\bibfnamefont {M.}~\bibnamefont
  {Gorghetto}}, \bibinfo {author} {\bibfnamefont {E.}~\bibnamefont {Hardy}}, \
  and\ \bibinfo {author} {\bibfnamefont {G.}~\bibnamefont {Villadoro}},\ }\href
  {\doibase 10.21468/SciPostPhys.10.2.050} {\bibfield  {journal} {\bibinfo
  {journal} {SciPost Phys.}\ }\textbf {\bibinfo {volume} {10}},\ \bibinfo
  {pages} {050} (\bibinfo {year} {2021}{\natexlab{a}})},\ \Eprint
  {http://arxiv.org/abs/2007.04990} {arXiv:2007.04990 [hep-ph]} \BibitemShut
  {NoStop}%
\bibitem [{\citenamefont {Buschmann}\ \emph {et~al.}(2022)\citenamefont
  {Buschmann}, \citenamefont {Foster}, \citenamefont {Hook}, \citenamefont
  {Peterson}, \citenamefont {Willcox}, \citenamefont {Zhang},\ and\
  \citenamefont {Safdi}}]{Buschmann_2022}%
  \BibitemOpen
  \bibfield  {author} {\bibinfo {author} {\bibfnamefont {M.}~\bibnamefont
  {Buschmann}}, \bibinfo {author} {\bibfnamefont {J.~W.}\ \bibnamefont
  {Foster}}, \bibinfo {author} {\bibfnamefont {A.}~\bibnamefont {Hook}},
  \bibinfo {author} {\bibfnamefont {A.}~\bibnamefont {Peterson}}, \bibinfo
  {author} {\bibfnamefont {D.~E.}\ \bibnamefont {Willcox}}, \bibinfo {author}
  {\bibfnamefont {W.}~\bibnamefont {Zhang}}, \ and\ \bibinfo {author}
  {\bibfnamefont {B.~R.}\ \bibnamefont {Safdi}},\ }\href {\doibase
  10.1038/s41467-022-28669-y} {\bibfield  {journal} {\bibinfo  {journal}
  {Nature Communications}\ }\textbf {\bibinfo {volume} {13}} (\bibinfo {year}
  {2022}),\ 10.1038/s41467-022-28669-y}\BibitemShut {NoStop}%
\bibitem [{\citenamefont {Benabou}\ \emph {et~al.}(2024)\citenamefont
  {Benabou}, \citenamefont {Buschmann}, \citenamefont {Kumar}, \citenamefont
  {Park},\ and\ \citenamefont {Safdi}}]{Benabou:2023ghl}%
  \BibitemOpen
  \bibfield  {author} {\bibinfo {author} {\bibfnamefont {J.~N.}\ \bibnamefont
  {Benabou}}, \bibinfo {author} {\bibfnamefont {M.}~\bibnamefont {Buschmann}},
  \bibinfo {author} {\bibfnamefont {S.}~\bibnamefont {Kumar}}, \bibinfo
  {author} {\bibfnamefont {Y.}~\bibnamefont {Park}}, \ and\ \bibinfo {author}
  {\bibfnamefont {B.~R.}\ \bibnamefont {Safdi}},\ }\href {\doibase
  10.1103/PhysRevD.109.055005} {\bibfield  {journal} {\bibinfo  {journal}
  {Phys. Rev. D}\ }\textbf {\bibinfo {volume} {109}},\ \bibinfo {pages}
  {055005} (\bibinfo {year} {2024})},\ \Eprint
  {http://arxiv.org/abs/2308.01334} {arXiv:2308.01334 [hep-ph]} \BibitemShut
  {NoStop}%
\bibitem [{\citenamefont {Benabou}\ \emph {et~al.}(2025)\citenamefont
  {Benabou}, \citenamefont {Buschmann}, \citenamefont {Foster},\ and\
  \citenamefont {Safdi}}]{Benabou:2024msj}%
  \BibitemOpen
  \bibfield  {author} {\bibinfo {author} {\bibfnamefont {J.~N.}\ \bibnamefont
  {Benabou}}, \bibinfo {author} {\bibfnamefont {M.}~\bibnamefont {Buschmann}},
  \bibinfo {author} {\bibfnamefont {J.~W.}\ \bibnamefont {Foster}}, \ and\
  \bibinfo {author} {\bibfnamefont {B.~R.}\ \bibnamefont {Safdi}},\ }\href
  {\doibase 10.1103/6v21-d6sj} {\bibfield  {journal} {\bibinfo  {journal}
  {Phys. Rev. Lett.}\ }\textbf {\bibinfo {volume} {134}},\ \bibinfo {pages}
  {241003} (\bibinfo {year} {2025})},\ \Eprint
  {http://arxiv.org/abs/2412.08699} {arXiv:2412.08699 [hep-ph]} \BibitemShut
  {NoStop}%
\bibitem [{\citenamefont {Saikawa}\ \emph {et~al.}(2024)\citenamefont
  {Saikawa}, \citenamefont {Redondo}, \citenamefont {Vaquero},\ and\
  \citenamefont {Kaltschmidt}}]{Saikawa:2024bta}%
  \BibitemOpen
  \bibfield  {author} {\bibinfo {author} {\bibfnamefont {K.}~\bibnamefont
  {Saikawa}}, \bibinfo {author} {\bibfnamefont {J.}~\bibnamefont {Redondo}},
  \bibinfo {author} {\bibfnamefont {A.}~\bibnamefont {Vaquero}}, \ and\
  \bibinfo {author} {\bibfnamefont {M.}~\bibnamefont {Kaltschmidt}},\
  }\href@noop {} {\  (\bibinfo {year} {2024})},\ \Eprint
  {http://arxiv.org/abs/2401.17253} {arXiv:2401.17253 [hep-ph]} \BibitemShut
  {NoStop}%
\bibitem [{\citenamefont {Kim}\ \emph {et~al.}(2024)\citenamefont {Kim},
  \citenamefont {Park},\ and\ \citenamefont {Son}}]{Kim:2024wku}%
  \BibitemOpen
  \bibfield  {author} {\bibinfo {author} {\bibfnamefont {H.}~\bibnamefont
  {Kim}}, \bibinfo {author} {\bibfnamefont {J.}~\bibnamefont {Park}}, \ and\
  \bibinfo {author} {\bibfnamefont {M.}~\bibnamefont {Son}},\ }\href {\doibase
  10.1007/JHEP07(2024)150} {\bibfield  {journal} {\bibinfo  {journal} {JHEP}\
  }\textbf {\bibinfo {volume} {07}},\ \bibinfo {pages} {150} (\bibinfo {year}
  {2024})},\ \Eprint {http://arxiv.org/abs/2402.00741} {arXiv:2402.00741
  [hep-ph]} \BibitemShut {NoStop}%
\bibitem [{\citenamefont {Baeza-Ballesteros}\ \emph {et~al.}(2024)\citenamefont
  {Baeza-Ballesteros}, \citenamefont {Copeland}, \citenamefont {Figueroa},\
  and\ \citenamefont {Lizarraga}}]{Baeza-Ballesteros:2023say}%
  \BibitemOpen
  \bibfield  {author} {\bibinfo {author} {\bibfnamefont {J.}~\bibnamefont
  {Baeza-Ballesteros}}, \bibinfo {author} {\bibfnamefont {E.~J.}\ \bibnamefont
  {Copeland}}, \bibinfo {author} {\bibfnamefont {D.~G.}\ \bibnamefont
  {Figueroa}}, \ and\ \bibinfo {author} {\bibfnamefont {J.}~\bibnamefont
  {Lizarraga}},\ }\href {\doibase 10.1103/PhysRevD.110.043522} {\bibfield
  {journal} {\bibinfo  {journal} {Phys. Rev. D}\ }\textbf {\bibinfo {volume}
  {110}},\ \bibinfo {pages} {043522} (\bibinfo {year} {2024})},\ \Eprint
  {http://arxiv.org/abs/2308.08456} {arXiv:2308.08456 [astro-ph.CO]}
  \BibitemShut {NoStop}%
\bibitem [{\citenamefont {Vachaspati}\ and\ \citenamefont
  {Vilenkin}(1985)}]{Vachaspati:1984gt}%
  \BibitemOpen
  \bibfield  {author} {\bibinfo {author} {\bibfnamefont {T.}~\bibnamefont
  {Vachaspati}}\ and\ \bibinfo {author} {\bibfnamefont {A.}~\bibnamefont
  {Vilenkin}},\ }\href {\doibase 10.1103/PhysRevD.31.3052} {\bibfield
  {journal} {\bibinfo  {journal} {Phys. Rev. D}\ }\textbf {\bibinfo {volume}
  {31}},\ \bibinfo {pages} {3052} (\bibinfo {year} {1985})}\BibitemShut
  {NoStop}%
\bibitem [{\citenamefont {Gorghetto}\ \emph
  {et~al.}(2021{\natexlab{b}})\citenamefont {Gorghetto}, \citenamefont
  {Hardy},\ and\ \citenamefont {Nicolaescu}}]{Gorghetto:2021fsn}%
  \BibitemOpen
  \bibfield  {author} {\bibinfo {author} {\bibfnamefont {M.}~\bibnamefont
  {Gorghetto}}, \bibinfo {author} {\bibfnamefont {E.}~\bibnamefont {Hardy}}, \
  and\ \bibinfo {author} {\bibfnamefont {H.}~\bibnamefont {Nicolaescu}},\
  }\href {\doibase 10.1088/1475-7516/2021/06/034} {\bibfield  {journal}
  {\bibinfo  {journal} {JCAP}\ }\textbf {\bibinfo {volume} {06}},\ \bibinfo
  {pages} {034} (\bibinfo {year} {2021}{\natexlab{b}})},\ \Eprint
  {http://arxiv.org/abs/2101.11007} {arXiv:2101.11007 [hep-ph]} \BibitemShut
  {NoStop}%
\bibitem [{\citenamefont {Zhao}\ \emph {et~al.}(2026)\citenamefont {Zhao},
  \citenamefont {Bian},\ and\ \citenamefont {Shu}}]{Zhao:2025kvm}%
  \BibitemOpen
  \bibfield  {author} {\bibinfo {author} {\bibfnamefont {Z.}~\bibnamefont
  {Zhao}}, \bibinfo {author} {\bibfnamefont {L.}~\bibnamefont {Bian}}, \ and\
  \bibinfo {author} {\bibfnamefont {J.}~\bibnamefont {Shu}},\ }\href {\doibase
  10.1103/52nd-mg9m} {\bibfield  {journal} {\bibinfo  {journal} {Phys. Rev. D}\
  }\textbf {\bibinfo {volume} {114}},\ \bibinfo {pages} {023539} (\bibinfo
  {year} {2026})},\ \Eprint {http://arxiv.org/abs/2507.00685} {arXiv:2507.00685
  [astro-ph.CO]} \BibitemShut {NoStop}%
\bibitem [{\citenamefont {Hindmarsh}\ and\ \citenamefont
  {Kibble}(1995)}]{Hindmarsh:1994re}%
  \BibitemOpen
  \bibfield  {author} {\bibinfo {author} {\bibfnamefont {M.~B.}\ \bibnamefont
  {Hindmarsh}}\ and\ \bibinfo {author} {\bibfnamefont {T.~W.~B.}\ \bibnamefont
  {Kibble}},\ }\href {\doibase 10.1088/0034-4885/58/5/001} {\bibfield
  {journal} {\bibinfo  {journal} {Rept. Prog. Phys.}\ }\textbf {\bibinfo
  {volume} {58}},\ \bibinfo {pages} {477} (\bibinfo {year} {1995})},\ \Eprint
  {http://arxiv.org/abs/hep-ph/9411342} {arXiv:hep-ph/9411342} \BibitemShut
  {NoStop}%
\bibitem [{\citenamefont {Kibble}(1980)}]{Kibble:1980mv}%
  \BibitemOpen
  \bibfield  {author} {\bibinfo {author} {\bibfnamefont {T.~W.~B.}\
  \bibnamefont {Kibble}},\ }\href {\doibase 10.1016/0370-1573(80)90091-5}
  {\bibfield  {journal} {\bibinfo  {journal} {Phys. Rept.}\ }\textbf {\bibinfo
  {volume} {67}},\ \bibinfo {pages} {183} (\bibinfo {year} {1980})}\BibitemShut
  {NoStop}%
\bibitem [{\citenamefont {Hindmarsh}\ \emph {et~al.}(1994)\citenamefont
  {Hindmarsh}, \citenamefont {Davis},\ and\ \citenamefont
  {Brandenberger}}]{Hindmarsh:1993av}%
  \BibitemOpen
  \bibfield  {author} {\bibinfo {author} {\bibfnamefont {M.}~\bibnamefont
  {Hindmarsh}}, \bibinfo {author} {\bibfnamefont {A.-C.}\ \bibnamefont
  {Davis}}, \ and\ \bibinfo {author} {\bibfnamefont {R.~H.}\ \bibnamefont
  {Brandenberger}},\ }\href {\doibase 10.1103/PhysRevD.49.1944} {\bibfield
  {journal} {\bibinfo  {journal} {Phys. Rev. D}\ }\textbf {\bibinfo {volume}
  {49}},\ \bibinfo {pages} {1944} (\bibinfo {year} {1994})},\ \Eprint
  {http://arxiv.org/abs/hep-ph/9307203} {arXiv:hep-ph/9307203} \BibitemShut
  {NoStop}%
\bibitem [{\citenamefont {Copeland}\ and\ \citenamefont
  {Saffin}(1996)}]{Copeland:1996jz}%
  \BibitemOpen
  \bibfield  {author} {\bibinfo {author} {\bibfnamefont {E.~J.}\ \bibnamefont
  {Copeland}}\ and\ \bibinfo {author} {\bibfnamefont {P.~M.}\ \bibnamefont
  {Saffin}},\ }\href {\doibase 10.1103/PhysRevD.54.6088} {\bibfield  {journal}
  {\bibinfo  {journal} {Phys. Rev. D}\ }\textbf {\bibinfo {volume} {54}},\
  \bibinfo {pages} {6088} (\bibinfo {year} {1996})},\ \Eprint
  {http://arxiv.org/abs/hep-ph/9604231} {arXiv:hep-ph/9604231} \BibitemShut
  {NoStop}%
\bibitem [{\citenamefont {Kibble}\ and\ \citenamefont
  {Vilenkin}(1995)}]{Kibble:1995aa}%
  \BibitemOpen
  \bibfield  {author} {\bibinfo {author} {\bibfnamefont {T.~W.~B.}\
  \bibnamefont {Kibble}}\ and\ \bibinfo {author} {\bibfnamefont
  {A.}~\bibnamefont {Vilenkin}},\ }\href {\doibase 10.1103/PhysRevD.52.679}
  {\bibfield  {journal} {\bibinfo  {journal} {Phys. Rev. D}\ }\textbf {\bibinfo
  {volume} {52}},\ \bibinfo {pages} {679} (\bibinfo {year} {1995})},\ \Eprint
  {http://arxiv.org/abs/hep-ph/9501266} {arXiv:hep-ph/9501266} \BibitemShut
  {NoStop}%
\bibitem [{\citenamefont {Srivastava}(1992)}]{Srivastava:1991nv}%
  \BibitemOpen
  \bibfield  {author} {\bibinfo {author} {\bibfnamefont {A.~M.}\ \bibnamefont
  {Srivastava}},\ }\href {\doibase 10.1103/PhysRevD.46.1353} {\bibfield
  {journal} {\bibinfo  {journal} {Phys. Rev. D}\ }\textbf {\bibinfo {volume}
  {46}},\ \bibinfo {pages} {1353} (\bibinfo {year} {1992})}\BibitemShut
  {NoStop}%
\bibitem [{\citenamefont {Melfo}\ and\ \citenamefont
  {Perivolaropoulos}(1995)}]{Melfo:1994cv}%
  \BibitemOpen
  \bibfield  {author} {\bibinfo {author} {\bibfnamefont {A.}~\bibnamefont
  {Melfo}}\ and\ \bibinfo {author} {\bibfnamefont {L.}~\bibnamefont
  {Perivolaropoulos}},\ }\href {\doibase 10.1103/PhysRevD.52.992} {\bibfield
  {journal} {\bibinfo  {journal} {Phys. Rev. D}\ }\textbf {\bibinfo {volume}
  {52}},\ \bibinfo {pages} {992} (\bibinfo {year} {1995})},\ \Eprint
  {http://arxiv.org/abs/hep-ph/9501284} {arXiv:hep-ph/9501284} \BibitemShut
  {NoStop}%
\bibitem [{\citenamefont {Ferrera}(1998)}]{Ferrera:1996hu}%
  \BibitemOpen
  \bibfield  {author} {\bibinfo {author} {\bibfnamefont {A.}~\bibnamefont
  {Ferrera}},\ }\href {\doibase 10.1103/PhysRevD.57.7130} {\bibfield  {journal}
  {\bibinfo  {journal} {Phys. Rev. D}\ }\textbf {\bibinfo {volume} {57}},\
  \bibinfo {pages} {7130} (\bibinfo {year} {1998})},\ \Eprint
  {http://arxiv.org/abs/hep-ph/9612487} {arXiv:hep-ph/9612487} \BibitemShut
  {NoStop}%
\bibitem [{\citenamefont {Lilley}\ and\ \citenamefont
  {Ferrera}(2001)}]{Lilley:2001df}%
  \BibitemOpen
  \bibfield  {author} {\bibinfo {author} {\bibfnamefont {M.}~\bibnamefont
  {Lilley}}\ and\ \bibinfo {author} {\bibfnamefont {A.}~\bibnamefont
  {Ferrera}},\ }\href {\doibase 10.1103/PhysRevD.64.023520} {\bibfield
  {journal} {\bibinfo  {journal} {Phys. Rev. D}\ }\textbf {\bibinfo {volume}
  {64}},\ \bibinfo {pages} {023520} (\bibinfo {year} {2001})},\ \Eprint
  {http://arxiv.org/abs/hep-ph/0102035} {arXiv:hep-ph/0102035} \BibitemShut
  {NoStop}%
\bibitem [{\citenamefont {Figueroa}\ \emph {et~al.}(2021)\citenamefont
  {Figueroa}, \citenamefont {Florio}, \citenamefont {Torrenti},\ and\
  \citenamefont {Valkenburg}}]{Figueroa:2021yhd}%
  \BibitemOpen
  \bibfield  {author} {\bibinfo {author} {\bibfnamefont {D.~G.}\ \bibnamefont
  {Figueroa}}, \bibinfo {author} {\bibfnamefont {A.}~\bibnamefont {Florio}},
  \bibinfo {author} {\bibfnamefont {F.}~\bibnamefont {Torrenti}}, \ and\
  \bibinfo {author} {\bibfnamefont {W.}~\bibnamefont {Valkenburg}},\ }\href
  {\doibase 10.1088/1475-7516/2021/04/035} {\bibfield  {journal} {\bibinfo
  {journal} {Journal of Cosmology and Astroparticle Physics}\ }\textbf
  {\bibinfo {volume} {2021}},\ \bibinfo {pages} {035} (\bibinfo {year}
  {2021})}\BibitemShut {NoStop}%
\bibitem [{\citenamefont {Figueroa}\ \emph {et~al.}(2023)\citenamefont
  {Figueroa}, \citenamefont {Florio}, \citenamefont {Torrenti},\ and\
  \citenamefont {Valkenburg}}]{Figueroa:2020rrl}%
  \BibitemOpen
  \bibfield  {author} {\bibinfo {author} {\bibfnamefont {D.~G.}\ \bibnamefont
  {Figueroa}}, \bibinfo {author} {\bibfnamefont {A.}~\bibnamefont {Florio}},
  \bibinfo {author} {\bibfnamefont {F.}~\bibnamefont {Torrenti}}, \ and\
  \bibinfo {author} {\bibfnamefont {W.}~\bibnamefont {Valkenburg}},\ }\href
  {\doibase 10.1016/j.cpc.2022.108586} {\bibfield  {journal} {\bibinfo
  {journal} {Computer Physics Communications}\ }\textbf {\bibinfo {volume}
  {283}},\ \bibinfo {pages} {108586} (\bibinfo {year} {2023})}\BibitemShut
  {NoStop}%
\bibitem [{\citenamefont {Guada}\ \emph {et~al.}(2019)\citenamefont {Guada},
  \citenamefont {Maiezza},\ and\ \citenamefont
  {Nemev{\v{s}}ek}}]{Guada:2018jek}%
  \BibitemOpen
  \bibfield  {author} {\bibinfo {author} {\bibfnamefont {V.}~\bibnamefont
  {Guada}}, \bibinfo {author} {\bibfnamefont {A.}~\bibnamefont {Maiezza}}, \
  and\ \bibinfo {author} {\bibfnamefont {M.}~\bibnamefont {Nemev{\v{s}}ek}},\
  }\href {\doibase 10.1103/PhysRevD.99.056020} {\bibfield  {journal} {\bibinfo
  {journal} {Phys. Rev. D}\ }\textbf {\bibinfo {volume} {99}},\ \bibinfo
  {pages} {056020} (\bibinfo {year} {2019})},\ \Eprint
  {http://arxiv.org/abs/1803.02227} {arXiv:1803.02227 [hep-th]} \BibitemShut
  {NoStop}%
\bibitem [{\citenamefont {Guada}\ \emph {et~al.}(2020)\citenamefont {Guada},
  \citenamefont {Nemev{\v{s}}ek},\ and\ \citenamefont
  {Pintar}}]{Guada:2020xnz}%
  \BibitemOpen
  \bibfield  {author} {\bibinfo {author} {\bibfnamefont {V.}~\bibnamefont
  {Guada}}, \bibinfo {author} {\bibfnamefont {M.}~\bibnamefont
  {Nemev{\v{s}}ek}}, \ and\ \bibinfo {author} {\bibfnamefont {M.}~\bibnamefont
  {Pintar}},\ }\href {\doibase 10.1016/j.cpc.2020.107480} {\bibfield  {journal}
  {\bibinfo  {journal} {Comput. Phys. Commun.}\ }\textbf {\bibinfo {volume}
  {256}},\ \bibinfo {pages} {107480} (\bibinfo {year} {2020})},\ \Eprint
  {http://arxiv.org/abs/2002.00881} {arXiv:2002.00881 [hep-ph]} \BibitemShut
  {NoStop}%
\bibitem [{\citenamefont {Hiramatsu}\ \emph {et~al.}(2011)\citenamefont
  {Hiramatsu}, \citenamefont {Kawasaki}, \citenamefont {Sekiguchi},
  \citenamefont {Yamaguchi},\ and\ \citenamefont {Yokoyama}}]{Hiramatsu_2011}%
  \BibitemOpen
  \bibfield  {author} {\bibinfo {author} {\bibfnamefont {T.}~\bibnamefont
  {Hiramatsu}}, \bibinfo {author} {\bibfnamefont {M.}~\bibnamefont {Kawasaki}},
  \bibinfo {author} {\bibfnamefont {T.}~\bibnamefont {Sekiguchi}}, \bibinfo
  {author} {\bibfnamefont {M.}~\bibnamefont {Yamaguchi}}, \ and\ \bibinfo
  {author} {\bibfnamefont {J.}~\bibnamefont {Yokoyama}},\ }\href {\doibase
  10.1103/physrevd.83.123531} {\bibfield  {journal} {\bibinfo  {journal}
  {Physical Review D}\ }\textbf {\bibinfo {volume} {83}} (\bibinfo {year}
  {2011}),\ 10.1103/physrevd.83.123531}\BibitemShut {NoStop}%
\bibitem [{\citenamefont {Jia}\ and\ \citenamefont {Bian}(2025)}]{Jia:2024zeu}%
  \BibitemOpen
  \bibfield  {author} {\bibinfo {author} {\bibfnamefont {Y.}~\bibnamefont
  {Jia}}\ and\ \bibinfo {author} {\bibfnamefont {L.}~\bibnamefont {Bian}},\
  }\href {\doibase 10.1103/PhysRevD.111.063552} {\bibfield  {journal} {\bibinfo
   {journal} {Phys. Rev. D}\ }\textbf {\bibinfo {volume} {111}},\ \bibinfo
  {pages} {063552} (\bibinfo {year} {2025})},\ \Eprint
  {http://arxiv.org/abs/2412.04218} {arXiv:2412.04218 [hep-ph]} \BibitemShut
  {NoStop}%
\bibitem [{\citenamefont {Amaro-Seoane}\ \emph {et~al.}(2017)\citenamefont
  {Amaro-Seoane} \emph {et~al.}}]{LISA:2017pwj}%
  \BibitemOpen
  \bibfield  {author} {\bibinfo {author} {\bibfnamefont {P.}~\bibnamefont
  {Amaro-Seoane}} \emph {et~al.} (\bibinfo {collaboration} {LISA}),\
  }\href@noop {} {\  (\bibinfo {year} {2017})},\ \Eprint
  {http://arxiv.org/abs/1702.00786} {arXiv:1702.00786 [astro-ph.IM]}
  \BibitemShut {NoStop}%
\bibitem [{\citenamefont {Baker}\ \emph {et~al.}(2019)\citenamefont {Baker}
  \emph {et~al.}}]{Baker:2019nia}%
  \BibitemOpen
  \bibfield  {author} {\bibinfo {author} {\bibfnamefont {J.}~\bibnamefont
  {Baker}} \emph {et~al.},\ }\href@noop {} {\  (\bibinfo {year} {2019})},\
  \Eprint {http://arxiv.org/abs/1907.06482} {arXiv:1907.06482 [astro-ph.IM]}
  \BibitemShut {NoStop}%
\bibitem [{\citenamefont {Luo}\ \emph {et~al.}(2016)\citenamefont {Luo} \emph
  {et~al.}}]{TianQin:2015yph}%
  \BibitemOpen
  \bibfield  {author} {\bibinfo {author} {\bibfnamefont {J.}~\bibnamefont
  {Luo}} \emph {et~al.} (\bibinfo {collaboration} {TianQin}),\ }\href {\doibase
  10.1088/0264-9381/33/3/035010} {\bibfield  {journal} {\bibinfo  {journal}
  {Class. Quant. Grav.}\ }\textbf {\bibinfo {volume} {33}},\ \bibinfo {pages}
  {035010} (\bibinfo {year} {2016})},\ \Eprint
  {http://arxiv.org/abs/1512.02076} {arXiv:1512.02076 [astro-ph.IM]}
  \BibitemShut {NoStop}%
\bibitem [{\citenamefont {Zhou}\ \emph {et~al.}(2023)\citenamefont {Zhou},
  \citenamefont {Cheng},\ and\ \citenamefont {Ren}}]{Zhou:2023rop}%
  \BibitemOpen
  \bibfield  {author} {\bibinfo {author} {\bibfnamefont {K.}~\bibnamefont
  {Zhou}}, \bibinfo {author} {\bibfnamefont {J.}~\bibnamefont {Cheng}}, \ and\
  \bibinfo {author} {\bibfnamefont {L.}~\bibnamefont {Ren}},\ }\href@noop {} {\
   (\bibinfo {year} {2023})},\ \Eprint {http://arxiv.org/abs/2306.14439}
  {arXiv:2306.14439 [gr-qc]} \BibitemShut {NoStop}%
\bibitem [{\citenamefont {Hu}\ and\ \citenamefont {Wu}(2017)}]{Hu:2017mde}%
  \BibitemOpen
  \bibfield  {author} {\bibinfo {author} {\bibfnamefont {W.-R.}\ \bibnamefont
  {Hu}}\ and\ \bibinfo {author} {\bibfnamefont {Y.-L.}\ \bibnamefont {Wu}},\
  }\href {\doibase 10.1093/nsr/nwx116} {\bibfield  {journal} {\bibinfo
  {journal} {Natl. Sci. Rev.}\ }\textbf {\bibinfo {volume} {4}},\ \bibinfo
  {pages} {685} (\bibinfo {year} {2017})}\BibitemShut {NoStop}%
\bibitem [{\citenamefont {Ruan}\ \emph {et~al.}(2020)\citenamefont {Ruan},
  \citenamefont {Guo}, \citenamefont {Cai},\ and\ \citenamefont
  {Zhang}}]{Ruan:2018tsw}%
  \BibitemOpen
  \bibfield  {author} {\bibinfo {author} {\bibfnamefont {W.-H.}\ \bibnamefont
  {Ruan}}, \bibinfo {author} {\bibfnamefont {Z.-K.}\ \bibnamefont {Guo}},
  \bibinfo {author} {\bibfnamefont {R.-G.}\ \bibnamefont {Cai}}, \ and\
  \bibinfo {author} {\bibfnamefont {Y.-Z.}\ \bibnamefont {Zhang}},\ }\href
  {\doibase 10.1142/S0217751X2050075X} {\bibfield  {journal} {\bibinfo
  {journal} {Int. J. Mod. Phys. A}\ }\textbf {\bibinfo {volume} {35}},\
  \bibinfo {pages} {2050075} (\bibinfo {year} {2020})},\ \Eprint
  {http://arxiv.org/abs/1807.09495} {arXiv:1807.09495 [gr-qc]} \BibitemShut
  {NoStop}%
\bibitem [{\citenamefont {Sesana}\ \emph {et~al.}(2021)\citenamefont {Sesana}
  \emph {et~al.}}]{Sesana:2019vho}%
  \BibitemOpen
  \bibfield  {author} {\bibinfo {author} {\bibfnamefont {A.}~\bibnamefont
  {Sesana}} \emph {et~al.},\ }\href {\doibase 10.1007/s10686-021-09709-9}
  {\bibfield  {journal} {\bibinfo  {journal} {Exper. Astron.}\ }\textbf
  {\bibinfo {volume} {51}},\ \bibinfo {pages} {1333} (\bibinfo {year}
  {2021})},\ \Eprint {http://arxiv.org/abs/1908.11391} {arXiv:1908.11391
  [astro-ph.IM]} \BibitemShut {NoStop}%
\bibitem [{\citenamefont {Seto}\ \emph {et~al.}(2001)\citenamefont {Seto},
  \citenamefont {Kawamura},\ and\ \citenamefont {Nakamura}}]{Seto:2001qf}%
  \BibitemOpen
  \bibfield  {author} {\bibinfo {author} {\bibfnamefont {N.}~\bibnamefont
  {Seto}}, \bibinfo {author} {\bibfnamefont {S.}~\bibnamefont {Kawamura}}, \
  and\ \bibinfo {author} {\bibfnamefont {T.}~\bibnamefont {Nakamura}},\ }\href
  {\doibase 10.1103/PhysRevLett.87.221103} {\bibfield  {journal} {\bibinfo
  {journal} {Phys. Rev. Lett.}\ }\textbf {\bibinfo {volume} {87}},\ \bibinfo
  {pages} {221103} (\bibinfo {year} {2001})},\ \Eprint
  {http://arxiv.org/abs/astro-ph/0108011} {arXiv:astro-ph/0108011} \BibitemShut
  {NoStop}%
\bibitem [{\citenamefont {Kawamura}\ \emph {et~al.}(2011)\citenamefont
  {Kawamura} \emph {et~al.}}]{Kawamura:2011zz}%
  \BibitemOpen
  \bibfield  {author} {\bibinfo {author} {\bibfnamefont {S.}~\bibnamefont
  {Kawamura}} \emph {et~al.},\ }\href {\doibase 10.1088/0264-9381/28/9/094011}
  {\bibfield  {journal} {\bibinfo  {journal} {Class. Quant. Grav.}\ }\textbf
  {\bibinfo {volume} {28}},\ \bibinfo {pages} {094011} (\bibinfo {year}
  {2011})}\BibitemShut {NoStop}%
\bibitem [{\citenamefont {Yagi}\ and\ \citenamefont
  {Seto}(2011)}]{Yagi:2011wg}%
  \BibitemOpen
  \bibfield  {author} {\bibinfo {author} {\bibfnamefont {K.}~\bibnamefont
  {Yagi}}\ and\ \bibinfo {author} {\bibfnamefont {N.}~\bibnamefont {Seto}},\
  }\href {\doibase 10.1103/PhysRevD.83.044011} {\bibfield  {journal} {\bibinfo
  {journal} {Phys. Rev. D}\ }\textbf {\bibinfo {volume} {83}},\ \bibinfo
  {pages} {044011} (\bibinfo {year} {2011})},\ \bibinfo {note} {[Erratum:
  Phys.Rev.D 95, 109901 (2017)]},\ \Eprint {http://arxiv.org/abs/1101.3940}
  {arXiv:1101.3940 [astro-ph.CO]} \BibitemShut {NoStop}%
\bibitem [{\citenamefont {Isoyama}\ \emph {et~al.}(2018)\citenamefont
  {Isoyama}, \citenamefont {Nakano},\ and\ \citenamefont
  {Nakamura}}]{Isoyama:2018rjb}%
  \BibitemOpen
  \bibfield  {author} {\bibinfo {author} {\bibfnamefont {S.}~\bibnamefont
  {Isoyama}}, \bibinfo {author} {\bibfnamefont {H.}~\bibnamefont {Nakano}}, \
  and\ \bibinfo {author} {\bibfnamefont {T.}~\bibnamefont {Nakamura}},\ }\href
  {\doibase 10.1093/ptep/pty078} {\bibfield  {journal} {\bibinfo  {journal}
  {PTEP}\ }\textbf {\bibinfo {volume} {2018}},\ \bibinfo {pages} {073E01}
  (\bibinfo {year} {2018})},\ \Eprint {http://arxiv.org/abs/1802.06977}
  {arXiv:1802.06977 [gr-qc]} \BibitemShut {NoStop}%
\bibitem [{\citenamefont {Crowder}\ and\ \citenamefont
  {Cornish}(2005)}]{Crowder:2005nr}%
  \BibitemOpen
  \bibfield  {author} {\bibinfo {author} {\bibfnamefont {J.}~\bibnamefont
  {Crowder}}\ and\ \bibinfo {author} {\bibfnamefont {N.~J.}\ \bibnamefont
  {Cornish}},\ }\href {\doibase 10.1103/PhysRevD.72.083005} {\bibfield
  {journal} {\bibinfo  {journal} {Phys. Rev. D}\ }\textbf {\bibinfo {volume}
  {72}},\ \bibinfo {pages} {083005} (\bibinfo {year} {2005})},\ \Eprint
  {http://arxiv.org/abs/gr-qc/0506015} {arXiv:gr-qc/0506015} \BibitemShut
  {NoStop}%
\bibitem [{\citenamefont {Corbin}\ and\ \citenamefont
  {Cornish}(2006)}]{Corbin:2005ny}%
  \BibitemOpen
  \bibfield  {author} {\bibinfo {author} {\bibfnamefont {V.}~\bibnamefont
  {Corbin}}\ and\ \bibinfo {author} {\bibfnamefont {N.~J.}\ \bibnamefont
  {Cornish}},\ }\href {\doibase 10.1088/0264-9381/23/7/014} {\bibfield
  {journal} {\bibinfo  {journal} {Class. Quant. Grav.}\ }\textbf {\bibinfo
  {volume} {23}},\ \bibinfo {pages} {2435} (\bibinfo {year} {2006})},\ \Eprint
  {http://arxiv.org/abs/gr-qc/0512039} {arXiv:gr-qc/0512039} \BibitemShut
  {NoStop}%
\bibitem [{\citenamefont {Harry}\ \emph {et~al.}(2006)\citenamefont {Harry},
  \citenamefont {Fritschel}, \citenamefont {Shaddock}, \citenamefont
  {Folkner},\ and\ \citenamefont {Phinney}}]{Harry:2006fi}%
  \BibitemOpen
  \bibfield  {author} {\bibinfo {author} {\bibfnamefont {G.~M.}\ \bibnamefont
  {Harry}}, \bibinfo {author} {\bibfnamefont {P.}~\bibnamefont {Fritschel}},
  \bibinfo {author} {\bibfnamefont {D.~A.}\ \bibnamefont {Shaddock}}, \bibinfo
  {author} {\bibfnamefont {W.}~\bibnamefont {Folkner}}, \ and\ \bibinfo
  {author} {\bibfnamefont {E.~S.}\ \bibnamefont {Phinney}},\ }\href {\doibase
  10.1088/0264-9381/23/15/008} {\bibfield  {journal} {\bibinfo  {journal}
  {Class. Quant. Grav.}\ }\textbf {\bibinfo {volume} {23}},\ \bibinfo {pages}
  {4887} (\bibinfo {year} {2006})},\ \bibinfo {note} {[Erratum:
  Class.Quant.Grav. 23, 7361 (2006)]}\BibitemShut {NoStop}%
\bibitem [{\citenamefont {Janssen}\ \emph {et~al.}(2015)\citenamefont {Janssen}
  \emph {et~al.}}]{Janssen:2014dka}%
  \BibitemOpen
  \bibfield  {author} {\bibinfo {author} {\bibfnamefont {G.}~\bibnamefont
  {Janssen}} \emph {et~al.},\ }\href {\doibase 10.22323/1.215.0037} {\bibfield
  {journal} {\bibinfo  {journal} {PoS}\ }\textbf {\bibinfo {volume}
  {AASKA14}},\ \bibinfo {pages} {037} (\bibinfo {year} {2015})},\ \Eprint
  {http://arxiv.org/abs/1501.00127} {arXiv:1501.00127 [astro-ph.IM]}
  \BibitemShut {NoStop}%
\bibitem [{\citenamefont {Punturo}\ \emph {et~al.}(2010)\citenamefont {Punturo}
  \emph {et~al.}}]{Punturo:2010zz}%
  \BibitemOpen
  \bibfield  {author} {\bibinfo {author} {\bibfnamefont {M.}~\bibnamefont
  {Punturo}} \emph {et~al.},\ }\href {\doibase 10.1088/0264-9381/27/19/194002}
  {\bibfield  {journal} {\bibinfo  {journal} {Class. Quant. Grav.}\ }\textbf
  {\bibinfo {volume} {27}},\ \bibinfo {pages} {194002} (\bibinfo {year}
  {2010})}\BibitemShut {NoStop}%
\bibitem [{\citenamefont {Hild}\ \emph {et~al.}(2011)\citenamefont {Hild} \emph
  {et~al.}}]{Hild:2010id}%
  \BibitemOpen
  \bibfield  {author} {\bibinfo {author} {\bibfnamefont {S.}~\bibnamefont
  {Hild}} \emph {et~al.},\ }\href {\doibase 10.1088/0264-9381/28/9/094013}
  {\bibfield  {journal} {\bibinfo  {journal} {Class. Quant. Grav.}\ }\textbf
  {\bibinfo {volume} {28}},\ \bibinfo {pages} {094013} (\bibinfo {year}
  {2011})},\ \Eprint {http://arxiv.org/abs/1012.0908} {arXiv:1012.0908 [gr-qc]}
  \BibitemShut {NoStop}%
\bibitem [{\citenamefont {Abbott}\ \emph {et~al.}(2017)\citenamefont {Abbott}
  \emph {et~al.}}]{LIGOScientific:2016wof}%
  \BibitemOpen
  \bibfield  {author} {\bibinfo {author} {\bibfnamefont {B.~P.}\ \bibnamefont
  {Abbott}} \emph {et~al.} (\bibinfo {collaboration} {LIGO Scientific}),\
  }\href {\doibase 10.1088/1361-6382/aa51f4} {\bibfield  {journal} {\bibinfo
  {journal} {Class. Quant. Grav.}\ }\textbf {\bibinfo {volume} {34}},\ \bibinfo
  {pages} {044001} (\bibinfo {year} {2017})},\ \Eprint
  {http://arxiv.org/abs/1607.08697} {arXiv:1607.08697 [astro-ph.IM]}
  \BibitemShut {NoStop}%
\bibitem [{\citenamefont {Aasi}\ \emph {et~al.}(2015)\citenamefont {Aasi} \emph
  {et~al.}}]{LIGOScientific:2014qfs}%
  \BibitemOpen
  \bibfield  {author} {\bibinfo {author} {\bibfnamefont {J.}~\bibnamefont
  {Aasi}} \emph {et~al.} (\bibinfo {collaboration} {LIGO Scientific, VIRGO}),\
  }\href {\doibase 10.1088/0264-9381/32/11/115012} {\bibfield  {journal}
  {\bibinfo  {journal} {Class. Quant. Grav.}\ }\textbf {\bibinfo {volume}
  {32}},\ \bibinfo {pages} {115012} (\bibinfo {year} {2015})},\ \Eprint
  {http://arxiv.org/abs/1410.7764} {arXiv:1410.7764 [gr-qc]} \BibitemShut
  {NoStop}%
\bibitem [{\citenamefont {Thrane}\ and\ \citenamefont
  {Romano}(2013)}]{Thrane:2013oya}%
  \BibitemOpen
  \bibfield  {author} {\bibinfo {author} {\bibfnamefont {E.}~\bibnamefont
  {Thrane}}\ and\ \bibinfo {author} {\bibfnamefont {J.~D.}\ \bibnamefont
  {Romano}},\ }\href {\doibase 10.1103/PhysRevD.88.124032} {\bibfield
  {journal} {\bibinfo  {journal} {Phys. Rev. D}\ }\textbf {\bibinfo {volume}
  {88}},\ \bibinfo {pages} {124032} (\bibinfo {year} {2013})},\ \Eprint
  {http://arxiv.org/abs/1310.5300} {arXiv:1310.5300 [astro-ph.IM]} \BibitemShut
  {NoStop}%
\bibitem [{\citenamefont {Abbott}\ \emph {et~al.}(2019)\citenamefont {Abbott}
  \emph {et~al.}}]{LIGOScientific:2019vic}%
  \BibitemOpen
  \bibfield  {author} {\bibinfo {author} {\bibfnamefont {B.~P.}\ \bibnamefont
  {Abbott}} \emph {et~al.} (\bibinfo {collaboration} {LIGO Scientific,
  Virgo}),\ }\href {\doibase 10.1103/PhysRevD.100.061101} {\bibfield  {journal}
  {\bibinfo  {journal} {Phys. Rev. D}\ }\textbf {\bibinfo {volume} {100}},\
  \bibinfo {pages} {061101} (\bibinfo {year} {2019})},\ \Eprint
  {http://arxiv.org/abs/1903.02886} {arXiv:1903.02886 [gr-qc]} \BibitemShut
  {NoStop}%
\bibitem [{\citenamefont {Ellis}\ \emph {et~al.}(2022)\citenamefont {Ellis},
  \citenamefont {Marsh}, \citenamefont {Eggemeier}, \citenamefont {Niemeyer},
  \citenamefont {Redondo},\ and\ \citenamefont {Dolag}}]{Ellis:2022grh}%
  \BibitemOpen
  \bibfield  {author} {\bibinfo {author} {\bibfnamefont {D.}~\bibnamefont
  {Ellis}}, \bibinfo {author} {\bibfnamefont {D.~J.~E.}\ \bibnamefont {Marsh}},
  \bibinfo {author} {\bibfnamefont {B.}~\bibnamefont {Eggemeier}}, \bibinfo
  {author} {\bibfnamefont {J.}~\bibnamefont {Niemeyer}}, \bibinfo {author}
  {\bibfnamefont {J.}~\bibnamefont {Redondo}}, \ and\ \bibinfo {author}
  {\bibfnamefont {K.}~\bibnamefont {Dolag}},\ }\href {\doibase
  10.1103/PhysRevD.106.103514} {\bibfield  {journal} {\bibinfo  {journal}
  {Phys. Rev. D}\ }\textbf {\bibinfo {volume} {106}},\ \bibinfo {pages}
  {103514} (\bibinfo {year} {2022})},\ \Eprint
  {http://arxiv.org/abs/2204.13187} {arXiv:2204.13187 [hep-ph]} \BibitemShut
  {NoStop}%
\bibitem [{\citenamefont {Eggemeier}\ \emph {et~al.}(2024)\citenamefont
  {Eggemeier}, \citenamefont {Anilkumar},\ and\ \citenamefont
  {Dolag}}]{Eggemeier:2024fzs}%
  \BibitemOpen
  \bibfield  {author} {\bibinfo {author} {\bibfnamefont {B.}~\bibnamefont
  {Eggemeier}}, \bibinfo {author} {\bibfnamefont {A.~K.}\ \bibnamefont
  {Anilkumar}}, \ and\ \bibinfo {author} {\bibfnamefont {K.}~\bibnamefont
  {Dolag}},\ }\href {\doibase 10.1103/PhysRevD.110.043530} {\bibfield
  {journal} {\bibinfo  {journal} {Phys. Rev. D}\ }\textbf {\bibinfo {volume}
  {110}},\ \bibinfo {pages} {043530} (\bibinfo {year} {2024})},\ \Eprint
  {http://arxiv.org/abs/2402.18221} {arXiv:2402.18221 [astro-ph.CO]}
  \BibitemShut {NoStop}%
\bibitem [{\citenamefont {Wantz}\ and\ \citenamefont
  {Shellard}(2010)}]{Wantz_2010}%
  \BibitemOpen
  \bibfield  {author} {\bibinfo {author} {\bibfnamefont {O.}~\bibnamefont
  {Wantz}}\ and\ \bibinfo {author} {\bibfnamefont {E.}~\bibnamefont
  {Shellard}},\ }\href {\doibase 10.1016/j.nuclphysb.2009.12.005} {\bibfield
  {journal} {\bibinfo  {journal} {Nuclear Physics B}\ }\textbf {\bibinfo
  {volume} {829}},\ \bibinfo {pages} {110–160} (\bibinfo {year}
  {2010})}\BibitemShut {NoStop}%
\bibitem [{\citenamefont {Kawasaki}\ \emph {et~al.}(2015)\citenamefont
  {Kawasaki}, \citenamefont {Saikawa},\ and\ \citenamefont
  {Sekiguchi}}]{PhysRevD.91.065014}%
  \BibitemOpen
  \bibfield  {author} {\bibinfo {author} {\bibfnamefont {M.}~\bibnamefont
  {Kawasaki}}, \bibinfo {author} {\bibfnamefont {K.}~\bibnamefont {Saikawa}}, \
  and\ \bibinfo {author} {\bibfnamefont {T.}~\bibnamefont {Sekiguchi}},\ }\href
  {\doibase 10.1103/PhysRevD.91.065014} {\bibfield  {journal} {\bibinfo
  {journal} {Phys. Rev. D}\ }\textbf {\bibinfo {volume} {91}},\ \bibinfo
  {pages} {065014} (\bibinfo {year} {2015})}\BibitemShut {NoStop}%
\end{thebibliography}%
\end{document}